\documentclass[
 reprint,
 amsmath,
 amssymb,
 aps, 
 groupedaddress, 
 pra
]{revtex4-2}

\usepackage{graphicx} % Required for inserting images
\usepackage{float}
\usepackage{dcolumn}  % Align table columns on decimal point
\usepackage{bm}       % Bold math
\usepackage[hidelinks]{hyperref}
\hypersetup{
    colorlinks=true,
    linkcolor=blue,
    urlcolor=blue,
    citecolor=blue,
}
\usepackage{soul} % for proper underline
\usepackage[table, dvipsnames]{xcolor}

\begin{document}

\title{Optical linewidth narrowing for device-coupled single T centers}
% \title{Optical linewidth narrowing for device-integrated single T centers}

\author{Adam Johnston\textsuperscript{1,2,$\dagger$}}
% \thanks{These authors contributed equally to this work}
\author{Yu-En Wong\textsuperscript{1,2,$\dagger$}}
% \thanks{These authors contributed equally to this work}
\author{Shengbin Yan\textsuperscript{3}}
\author{Ulises Felix-Rendon\textsuperscript{1,2}} 
\author{Songtao Chen\textsuperscript{1,4}}\thanks{songtao.chen@rice.edu}

% % \email{songtao.chen@rice.edu}
\affiliation{\textsuperscript{1}Department of Electrical and Computer Engineering, Rice University, Houston, TX 77005, USA}
\affiliation{\textsuperscript{2}Applied Physics Graduate Program, Smalley-Curl Institute, Rice University, Houston, TX 77005, USA}
\affiliation{\textsuperscript{3}Department of Physics and Astronomy, Rice University, Houston, TX 77005, USA}
\affiliation{\textsuperscript{4}Smalley-Curl Institute, Rice University, Houston, TX 77005, USA}
\thanks{These authors contributed equally to this work.}

% \author{Adam Johnston}
%     \thanks{These authors contributed equally to this work.}
%     \affiliation{Department of Electrical and Computer Engineering, Rice University, Houston, TX 77005, USA}
%     \affiliation{Applied Physics Graduate Program, Smalley-Curl Institute, Rice University, Houston, TX 77005, USA}
% \author{Yu-En Wong}
%     \thanks{These authors contributed equally to this work.}
%     \affiliation{Department of Electrical and Computer Engineering, Rice University, Houston, TX 77005, USA}
%     \affiliation{Applied Physics Graduate Program, Smalley-Curl Institute, Rice University, Houston, TX 77005, USA}
    
% \author{Shengbin Yan}
%     \affiliation{Department of Physics and Astronomy, Rice University, Houston, TX 77005, USA}
% \author{Ulises Felix}
%     \affiliation{Department of Electrical and Computer Engineering, Rice University, Houston, TX 77005, USA}
%     \affiliation{Applied Physics Graduate Program, Smalley-Curl Institute, Rice University, Houston, TX 77005, USA}
% \author{Songtao Chen}
%     \email{songtao.chen@rice.edu}
%     \affiliation{Department of Electrical and Computer Engineering, Rice University, Houston, TX 77005, USA}
%     \affiliation{Smalley-Curl Institute, Rice University, Houston, TX 77005, USA}

% \date{\today}

\begin{abstract}
Single T centers in silicon have emerged as promising optically active spins for quantum networking applications. One of the major obstacles to advancing the system is their broad optical linewidth due to spectral diffusion, which is two orders of magnitude larger than their cavity-enhanced radiative linewidth. We tackle this issue by utilizing above-band optical excitation
% at $\lambda = 980$ nm 
delivered via a laser scanning microscope to device-coupled single T centers, achieving up to 70\% optical linewidth reduction. We attribute the linewidth narrowing effect to the filling of nearby charge traps by photo-generated free carriers. We analyze charge stabilization dynamics by exploiting pulsed above-band excitation 
% revealing the lower bound of detrapping process larger than 300 $\mu$s in the dark. 
and develop a rate equation model to describe the dynamics and to explain the observed linewidth narrowing and center shift. This work provides an effective pathway to control and reduce the optical linewidth for single T centers, clearing one of the major roadblocks to advance the single T center spin platform for quantum information and networking applications.
\end{abstract}

\maketitle

\section{Introduction}

Optically interfaced solid-state spins \cite{awschalom2018quantum} are promising candidates for realizing quantum repeaters \cite{briegel1998quantum} aimed at constructing large-scale quantum networks -- a foundational technology for secure communication, distributed quantum computing and sensing \cite{awschalom2021development}. 
% Established platforms, such as nitrogen-vacancy (NV) and silicon-vacancy (SiV) centers in diamond as well as semiconductor quantum dots, have already achieved major milestones, including spin-photon entanglement \cite{togan2010quantum, gao2012observation, de2012quantum, nguyen2019integrated}, remote spin entanglement \cite{bernien2013heralded, knaut2024entanglement}, and memory-enhanced communication \cite{bhaskar2020experimental}. However, a primary challenge in harnessing these platforms for long-distance applications is that their native atomic transitions fall outside the low-loss fiber transmission window, necessitating complex nonlinear frequency conversion to telecom band \cite{de2012quantum, knaut2024entanglement, li2016efficient}. Conversely, single erbium ions (Er$^{3+}$) in oxide substrates provide direct transitions within the telecom C-band, enabling indistinguishable photon generation \cite{ourari2023indistinguishable} and spin-photon entanglement \cite{uysal2025spin} in hosts like CaWO$_4$. Nonetheless, achieving the strong evanescent coupling to the silicon photonic cavities integrated on the oxide surface requires Er$^{3+}$ ions to reside close to the surface, which can degrade their spin coherence due to the surface paramagnetic impurities \cite{ourari2023indistinguishable}.
% write lifetime vs coherence
%other spins in silicon Se, G, C, Er
Optically active spins in silicon (OASIS) are an emerging solid-state spin platform for quantum networking \cite{khoury2022bright, sandholzer2026single}, uniquely leveraging the inherent scalability enabled by the mature silicon manufacturing. Furthermore, isotopic enrichment can provide a magnetically quiet environment, creating a pristine ``semiconductor vacuum'' for the spins \cite{steger2012quantum}.
Among OASIS, the T center stands out due to its telecom O-band zero phonon line (ZPL) optical transition, a ground-state doublet spin manifold, long spin coherence times, and direct access to the $^1$H nuclear spin memory \cite{bergeron2020silicon}. Recent advancements have enabled the isolation of single T centers via nanophotonic integration \cite{higginbottom2022optical, islam2023cavity, johnston2024cavity} as well as the realization of a multi-qubit quantum register utilizing the T center electronic spin alongside $^1$H and $^{29}$Si nuclear spins \cite{song2026entanglement}. However, early attempts at remote spin entanglement achieved only modest fidelity and mHz entangling rate \cite{afzal2024distributed}, partially hindered by the broad optical linewidth of device-coupled single T centers, which is one of the major obstacles for advancing the T center platform.

Spectral diffusion (SD) remains a prevalent challenge for solid-state quantum emitters, severely limiting their optical coherence \cite{wolfowicz2021quantum}. Various strategies have been developed to mitigate or suppress this effect. High-purity type-IIa diamond has been used to generate NV centers exhibiting transform-limited linewidths \cite{tamarat2006stark}. In addition, leveraging inversion-symmetric and nonpolar lattice sites eliminates the permanent electric dipole to the first order, yielding near-lifetime-limited linewidths for Group-IV color centers in diamond \cite{zuber2023shallow, bhaskar2017quantum, trusheim2020transform, wang2024transform} and Er$^{3+}$ ions in CaWO$_4$ \cite{ourari2023indistinguishable}, respectively. Alternatively, embedding quantum defects within $p$-$i$-$n$ structures and utilizing electric fields to deplete and stabilize local charges have successfully mitigated spectral diffusion in VV$^0$ defect in SiC \cite{anderson2019electrical} and GaAs quantum dots \cite{zhai2020low}. Moreover, above-band laser excitation can be deployed to generate free carriers that fill charge traps, effectively minimizing charge noise \cite{majumdar2011effect, alizadehherfati2025electrical}. 

\begin{figure*}[t!] % The asterisk makes it double-column
    \centering
    \includegraphics[scale=1]{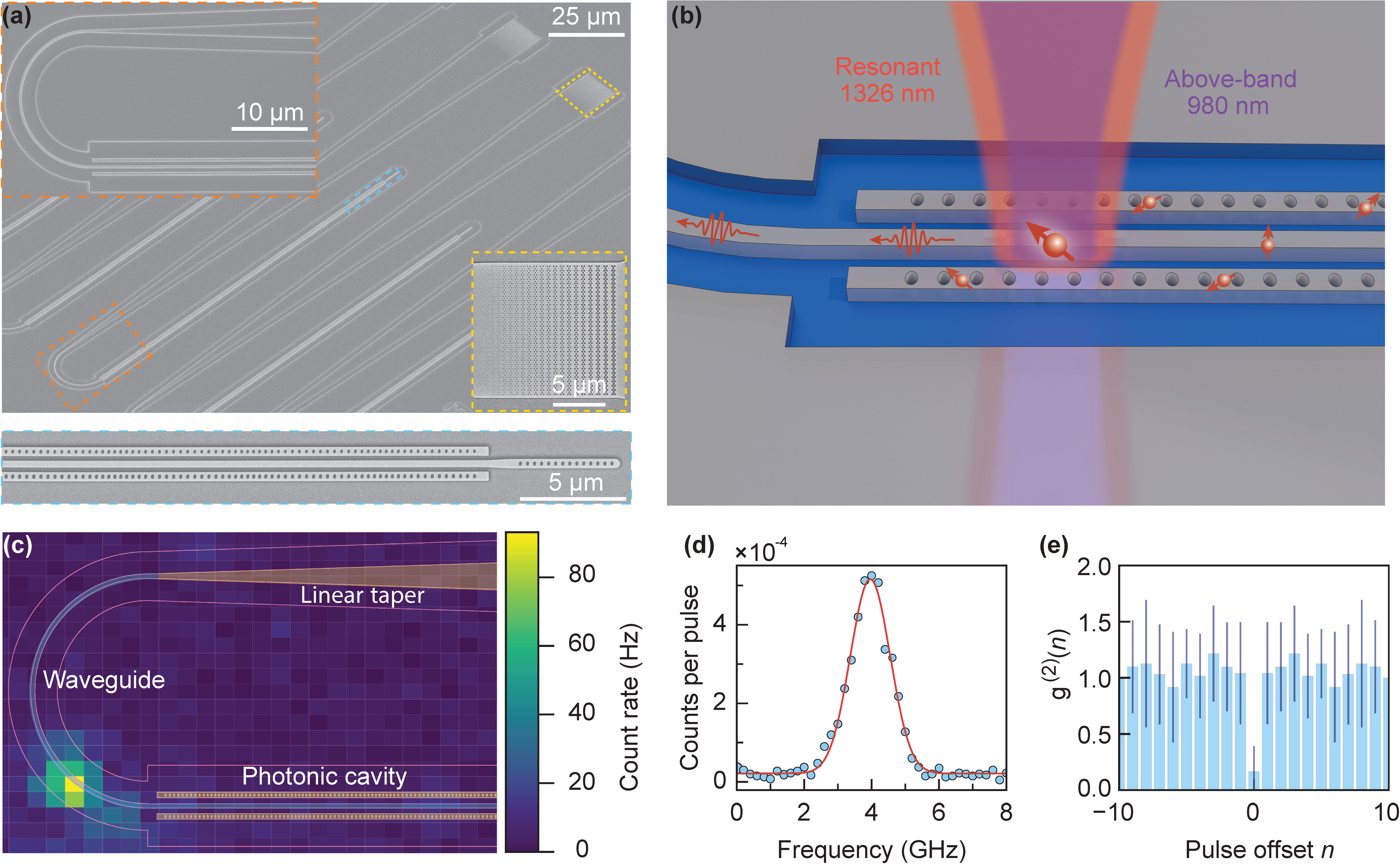}
    \caption{
    \textbf{Spatially-resolved PLE spectroscopy for device-coupled single T centers.}
    \textbf{(a)} Scanning electron microscope image of a block of five devices. Each device consists of two linear arrays of PC cavities (24 in total) evanescently coupled to a bus waveguide terminated with a broadband mirror (blue dashed rectangle), and a linear taper waveguide which connects  the GC (yellow dashed square) and the bus waveguide.
    \textbf{(b)} Schematics of the experiment. We use the resonant laser for PLE spectroscopy, and the above-band laser for controlling local charge environment. The two lasers are combined using a LSM setup. The T center fluorescence is collected via the waveguide and the GC to the SNSPD.
    \textbf{(c)} Spatial excitation scan with near resonant laser frequency. A single T center (named ``Itokawa'') is identified in the bus waveguide. Each pixel is $1\times1$ $\mu$m$^2$. Overlay of device geometry relies on the reflection scan using LSM
    % and PLE signal from a spatial scan at a fixed frequency. The color bar is photon count rate reading from SNSPD.
    \textbf{(d)} PLE spectrum of Itokawa. All frequencies in this paper are offset from $f_0 = 226130.674$ GHz. The red line shows the Gaussian fitting, revealing a FWHM linewidth of 1.38 $\pm$ 0.03 GHz.
    \textbf{(e)} Second-order autocorrelation measurement for Itokawa. All the photons detected after each excitation pulse are binned into a single time-bin. The horizontal axis shows the time-bin offset in unit of pulse sequence period (8~$\mu$s).
    }
    \label{fig1: PLE}
\end{figure*}

For T center ensembles in the highly purified, isotopically enriched bulk silicon ($^{28}$Si), a narrow inhomogeneous linewidth $\Gamma_\text{inh} = 33$ MHz has been observed \cite{bergeron2020silicon}. In the same substrate, the homogeneous linewidth measured via spectral hole burning reaches $\Gamma_\text{hom} = 0.69$ MHz \cite{deabreu2023waveguide}, which is only four times larger than the fundamental lifetime-limited linewidth of 0.17 MHz \cite{bergeron2020silicon}. However, nanophotonic device integration severely degrades this performance, broadening the T-center spectral linewidth to the $\sim$ GHz level \cite{islam2023cavity, johnston2024cavity, komza2025multiplexed}, which is about 200$\times$ larger than the Purcell-enhanced radiative linewidth (Appendix~\ref{appendix:lifetimeLimitedLinewidth}). This broadening has contributions from laser-induced spectral diffusion caused by the scrambling of the local charge environment, demonstrated by check-probe spectroscopy \cite{zhang2025laser,bowness2025laser}. Although implementing electric field control via $p$-$i$-$n$ structures successfully induces a DC Stark shift \cite{dobinson2025electrically, day2025probing, dobinson2026spectral}, it has yet to yield any optical linewidth narrowing.

Here, we report the successful realization of optical linewidth narrowing in device-coupled single T centers utilizing above-band optical excitation at $\lambda_\text{AB} = 980$ nm. By integrating this above-band excitation~into~a laser scanning microscope (LSM) setup alongside our time-resolved resonant photoluminescence excitation (PLE) spectroscopy (Appendix~\ref{appendix:detailedSetup}), we achieve up to 70\% linewidth narrowing. Furthermore, by probing the time dynamics under pulsed above-band excitation, we demonstrate that the narrowing can persist for more than 300 $\mu$s in the dark. We show that the linewidth narrowing effect strictly requires above-band photon energy by systematically sweeping $\lambda_{\text{AB}}$ across the silicon band gap. This dramatic improvement is attributed to photo-generated free carriers filling and stabilizing the local charge traps surrounding the single T centers \cite{majumdar2011effect, alizadehherfati2025electrical}, a mechanism validated by a rate-equation model. The model also well explains the observed center shift alongside the linewidth narrowing. Our work addresses a critical, long-standing bottleneck for single T centers, paving the way to advance this platform for next-generation quantum information and networking applications.

\begin{figure*}[t]
    \centering
    \includegraphics[scale=1]{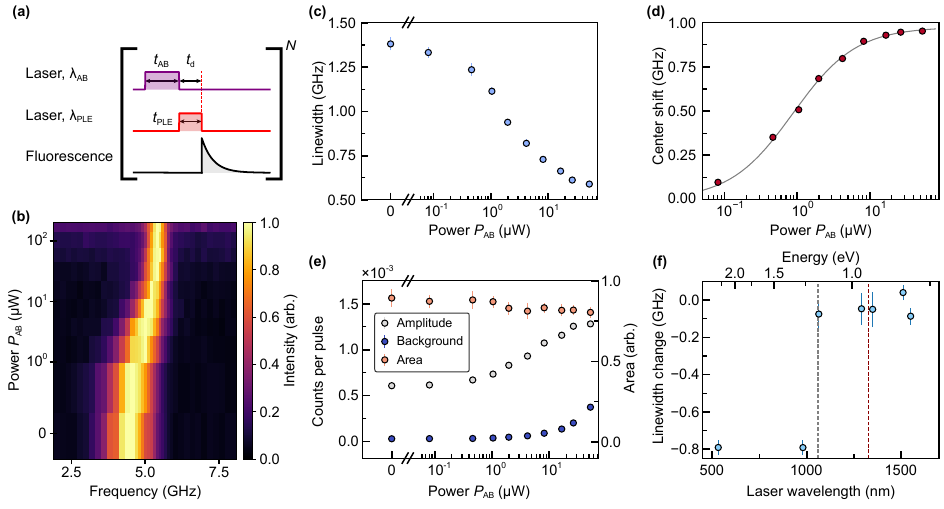}
    \caption{
    \textbf{T center linewidth narrowing via above-band excitation.}
    \textbf{(a)} Experimental pulse sequence. $t_\text{AB}$ and $t_\text{PLE}$ are the pulse widths of the above-band laser (at $\lambda_\text{AB}$) and the resonant laser (at $\lambda_\text{PLE}$), respectively; $t_\text{d}$ is the pulse separation and $N$ is the sequence repetition. For panels \textbf{(b}-\textbf{e)}, $t_\text{AB} = t_\text{PLE} = 500$ ns and $t_\text{d} = 0$ ns. 
    \textbf{(b)} Spectral linewidth narrowing for Itokawa under different above-band excitation power ($P_\text{AB}$). Panels \textbf{(c)}, \textbf{(d)}, and \textbf{(e)} describe the characteristics of Itokawa spectrum with varying $P_\text{AB}$. 
    \textbf{(c)} Gaussian-fitted FWHM linewidth.  
    \textbf{(d)} Center frequency shift against the no above-band excitation case. 
    The solid gray line is a fitted saturation curve with $P_\text{sat} = 0.87$ $\mu$W and an amplitude of 0.97 GHz.
    \textbf{(e)} Gaussian-fitted amplitude, background, and integrated area (background subtracted) of Itokawa spectra.
    \textbf{(f)} Maximum linewidth change of Itokawa under varying $\lambda_\text{AB}$, revealing a sharp cutoff in the linewidth-narrowing effect at the silicon bandgap (gray dashed line, $E_g = 1.17$ eV for $T$ $\approx$ 4 K). The red dashed line marks the position of the T center ZPL transition.
    %\textbf{(f)} A survey of the maximum observed frequency shift and linewidth change for 28 device-coupled single T centers under above-band excitation ($\lambda_\text{AB} = 980$ nm, $t_\text{AB} = 1000$ ns, $t_\text{PLE} = 500$ ns).
    The resonant excitation power is kept constant at  round 22 $\mu$W for all results in this paper. In all plots, error bars denote $\pm 1\sigma$ statistical uncertainty.
    % The error bar through out the whole paper represents the standard deviation for fitted values or propagated uncertainty from fitted values.  
    % OD0.0, 1326nm, 318*0.069 = 22 uW
    % this is power for 0507_Itokoawa dataset. Powers in f. vary < 1 uW. 
    }
    \label{fig2:PL_response}
\end{figure*}

% \section{Results}
\section{Spatially-resolved PLE spectroscopy}

% \subsection{Experiment Configuration}
Our nanophotonic devices are fabricated on a silicon-on-insulator (SOI) sample which is situated on the cold finger of a closed-cycle cryostat (\textit{T} = 3.6 K).  The thickness of the silicon device layer is 220 nm.
% and buried oxide layer are 220 nm and 2.3 $\mu$m, respectively. 
The T centers are generated via ion implantation followed by thermal annealing
% generation process is similar to that in our previous paper
\cite{johnston2024cavity}. Linear arrays of 1D photonic crystal (PC) cavities are evanescently coupled to a bus waveguide (Appendix~\ref{appendix:meanderDesign}), which is connected to a subwavelength grating coupler (GC) via a linearly tapered waveguide [Fig.~\ref{fig1: PLE}(a)]. An angle-polished fiber couples to the GC. %with a one-way coupling efficiency $\eta_\text{GC} =55.6$\% \cite{johnston2024cavity}.%
% GC measurement; 20260603
% The one-way GC coupling efficiency $\eta_\text{GC} = XX$\%, and the total system detection efficiency is XX\% \cite{johnston2024cavity}, which represents the probability for a fluorescence photon reached the GC and gets registered by the single photon detector. 
A LSM combines both resonant ($\lambda_\text{PLE}$ = 1326 nm) and above-band ($\lambda_\text{AB}$ = 980 nm) laser excitation, shining on single T centers through a 50$\times$ objective [Fig.~\ref{fig1: PLE}(b)]. The LSM features a close to diffraction-limited focused spot size of 1.7 $\mu$m at $\lambda_\text{PLE}$ = 1326 nm (Appendix~\ref{appendix:LSMSpotSize}). We note that, due to the narrow waveguide beam width (313 nm) and small absorption ($\sim 10^{-4}$) of $\lambda_\text{AB}$ = 980 nm photons in the device layer, the above-band laser pulses create rather uniform illumination in the waveguide width and depth directions. The resonant and above-band excitations are utilized for obtaining T center PLE spectrum and local charge environment stabilization, respectively. 
Fluorescence photons from the T center are coupled to the angle-polished fiber via the waveguide and the GC, then detected by a superconducting nanowire single photon detector (SNSPD). The detector is time gated to reject resonant excitation laser photons \cite{johnston2024cavity}, and spectral filters are used to minimize scattered photons at $\lambda_\text{AB}$ as well as background fluorescence beyond the T center~ZPL.
% The resonant laser pulse is used for PLE spectrum, and above band laser pulse is for modifying the local charge environment \cite{majumdar2011effect, anderson2019electrical, alizadehherfati2025electrical}.
% Emitted photons are optically coupled to the fiber via the waveguide and GC, then redirected to a SNSPD in another cryostat with \textit{T} = 2.7 K for detection. 

By controlling the resonant excitation frequency ($f_\text{exc}$) and scanning the focused laser spot, we can locate single T centers both spatially and spectrally with the LSM setup [Fig.~\ref{fig1: PLE}(c)]. 
% Photons emitted into the waveguide are coupled through the GC into an angle polished optical fiber \cite{johnston2024cavity}. A single photon detector in a separate cryostat registers the photon counts. 
% A single bright spot emerges when we perform a spatial scan with $f+\text{PLE} = 226134.764$ GHz [Fig.~\ref{fig1: PLE}(c)]. 
After spatially locating a single T center, we scan the laser frequency $f_\text{exc}$ to obtain the PLE spectrum [Fig.~\ref{fig1: PLE}(d)], revealing a full width at half maximum (FWHM) linewidth of $\Gamma = 1.38 \pm 0.03$ GHz. We note that iterative spatial and PLE scans are necessary to obtain an accurate location and spectral position for single T centers.
To verify that the peak originates from a single T center, we measure the second-order autocorrelation function $g^{(2)}$ using all the detected fluorescence photons after each excitation pulse [Fig.~\ref{fig1: PLE}(e)]. Photon antibunching is observed with the value $g^{(2)}$(0) = 0.17 $\pm$ 0.22,
% $g^{(2)}$(0) = 0.1662 $\pm$ 0.2249, 
which confirms that the majority of the detected photons come from a single emitter. For the remainder of this paper, we will refer to this single T center as ``Itokawa''.

\begin{figure*}[t]
    \centering
    \includegraphics[scale=1]{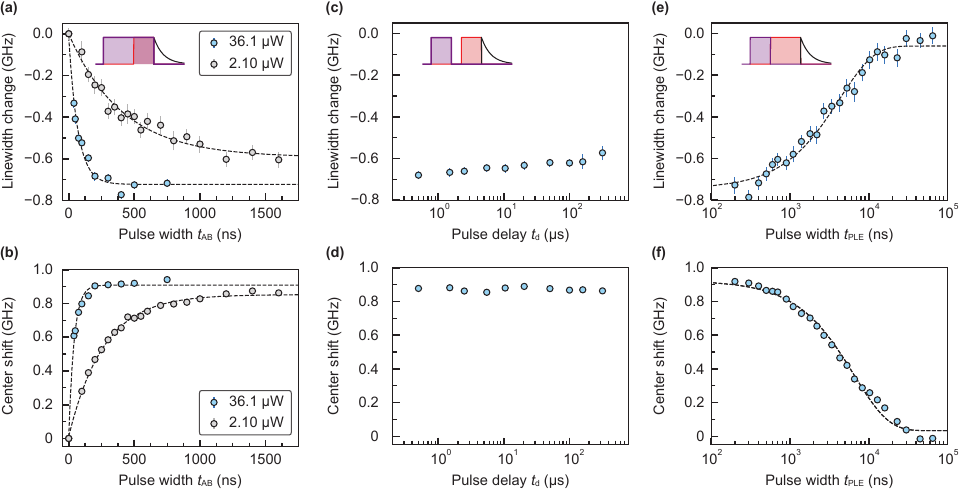}
    \caption{\textbf{Characteristic time scale for charge fluctuation dynamics.} 
        \textbf{(a)} Linewidth change and \textbf{(b)} center frequency shift of Itokawa spectrum under varying $t_\text{AB}$ and $P_\text{AB}$. The pulse delay and PLE pulse width remain same: $t_\text{d}=0$ ns and $t_\text{PLE}=500$~ns. The black dashed lines show the single exponential fitting, revealing time constants of $70\pm6.3$  and $41\pm2.6$ ns at $P_\text{AB} = 36.1 $~$\mu$W, $409\pm47$ and $266\pm8.5$ ns at $P_\text{AB} = 2.1$ $\mu$W, for linewidth change and center shift, respectively.
        \textbf{(c)} and \textbf{(d)} show the linewidth change and center frequency shift effect sustain more than $t_\text{d} = 300$ $\mu$s delay. The pulse widths $t_\text{AB} = t_\text{PLE} = 500$ ns.
        \textbf{(e)} and \textbf{(f)} shows the linewidth change and center frequency shift effect diminished by the 1326 nm resonant laser, with the delay $t_\text{d} = t_\text{PLE}$. The exponential fitting reveals the time constant of $4.3 \pm 0.4$ $\mu$s and $6.6 \pm 0.4$ $\mu$s, respectively.  The above-band laser pulse width and power are $t_\text{AB}= 500$ ns and $P_\text{AB} = $ 53.1 $\mu$W at $\lambda_\text{AB}=980 $ nm if not specified.
        % 168*0.2146~36.1
        % 14.9*0.2368 = 3.53 % we use interpolated value
        % 187.6*0.2654 = 49.8 uW
        % 318*0.0692 =22
        % 4298 and 6575 ns respectively.    
        % For \textbf{b-e}, the above-band laser pulsewidth is 500 ns, and the above-band laser power is chosen to be around the saturated power as shown in Fig.\ref{fig2:PL_response}. The black dashed lines are fitted exponential decay function.
        All the results are derived from Gaussian-fitted PLE spectrum.}
    \label{fig3:TimeDynamic}
\end{figure*}

% \subsection{Above-band Stimulation}
\section{Linewidth reduction via above-band optical excitation}\label{subsection:linewithReduction}

To investigate the T center linewidth reduction by the above-band optical excitation, we align above-band laser pulses within the time-resolved PLE spectroscopy pulse sequence [Fig.~\ref{fig2:PL_response}(a)] via TTL-controlled acousto-optic modulators.
With aligned above-band and resonant excitation pulses ($t_\text{AB} = t_\text{PLE} = 500$ ns and $t_\text{d} = 0$ ns), we characterize the response of the Itokawa spectrum with varying power $P_\text{AB}$ [Fig.~\ref{fig2:PL_response}(b)], revealing a decrease in the linewidth from 1.38 $\pm$ 0.03 GHz $(P_\text{AB} = 0$ $\mu$W) to 0.59 $\pm$ 0.01 GHz ($P_\text{AB} = 53.1$ $\mu$W) [Fig.~\ref{fig2:PL_response}(c)], which is accompanied by a center frequency blue shift of 0.95 $\pm$ 0.01 GHz [Fig.~\ref{fig2:PL_response}(d)]. The frequency shift is well fitted by a saturation curve, $\Delta \nu(P_\text{AB}) = \Delta \nu_{\infty}/(1 + P^\text{sat}_\text{AB}/P_\text{AB})$, revealing a saturation power of $P^\text{sat}_\text{AB} = 0.87$ $\mu$W. We note that the blue shift of the center frequency is opposite to the trend caused by elevated temperature, which helps us rule out the thermal effect to be the dominant factor. The above-band laser pulses do not contribute significantly to excite Itokawa, as shown by the minor change in the integrated area of its PLE spectrum under high $P_\text{AB}$ [Fig.~\ref{fig2:PL_response}(e)]. However, high $P_\text{AB}$ can lead to background increase [Fig.~\ref{fig2:PL_response}(e)], which is due to the excitation of background fluorescence in silicon that falls into the band-pass window of the spectral filters. 
% Figure \ref{fig2:PL_response}b-e show the pulse sequence and spectral response of the single T center.
% The T center response to 980 nm stimulation during PLE spectroscopy is characterized by a narrowing of the ZPL transition (Fig. \ref{fig2:PL_response}b) and a blue-shift of the resonant frequency (Fig. \ref{fig2:PL_response}c) as the above-band power increases. 

We attribute the linewidth narrowing effect to the charge environment stabilization due to the free carrier induced trap filling and neutralization \cite{majumdar2011effect, alizadehherfati2025electrical}. Higher above-band excitation power $P_\text{AB}$ generates more free carriers filling the charge traps, reducing the electric field noise for Itokawa. To verify the effect results from trap filling rather than ionization, we sweep $\lambda_\text{AB}$ across the silicon bandgap. We show that the narrowing effect strictly requires photon energy larger than the silicon bandgap [Fig.~\ref{fig2:PL_response}(f)], which indicates the participation of free carriers in the process. At $\lambda_\text{AB} = 532$ nm, similar linewidth change and center shift (Appendix ~\ref{appendix:532nm_response}) are observed as those measured under $\lambda_\text{AB} = 980$ nm.

% There is no significant dependence between the location of a T center and its linewidth change or frequency shift. However, the behavior of the outlying T centers suggest a connection between the magnitude of the frequency shift and linewidth change.

\section{Time dynamics of the linewidth narrowing effect}\label{subsection:TimeDynamics}

Next, we turn to the investigation of the time dynamics of the linewidth narrowing effect for Itokawa by controlling the resonant and above-band excitation pulse widths ($t_\text{PLE}$, $t_\text{AB}$), as well as the delay ($t_\text{d}$) between them [Fig.~\ref{fig2:PL_response}(a)]. To illustrate the trap filling process, we first vary above-band laser pulse width ($t_\text{AB}$) with the falling edge aligned to the resonant laser pulse, whose pulse width is kept at $t_\text{AB} = 500$ ns. Under a far-saturated power $P_\text{AB} =36.1 \,\mu$W, both linewidth narrowing [Fig.~\ref{fig3:TimeDynamic}(a)] and center shift [Fig.~\ref{fig3:TimeDynamic}(b)] exhibit single exponential decays with a time constant of $41 \pm 2.6$ ns and $70 \pm 6.3$ ns, respectively. At a lower power ($P_\text{AB} =2.10$ $\mu$W), the dynamics slow down to $266 \pm 8.5$ ns and $409 \pm 47$ ns for the above two cases, respectively. The higher above-band laser power promotes more rapid trap filling and neutralizing, which leads to faster linewidth narrowing and center shift changes.

We further study the stability of the charge environment in the dark. When increasing the pulse delay time $t_\text{d}$,
% between above-band and resonant lasers $(t_\text{d})$, 
we observe a very slow decay in the linewidth narrowing and center shift effect [Fig.~\ref{fig3:TimeDynamic}(c)(d)]. The pulse delay time is limited to 300 $\mu$s due to technical difficulties. The results suggest that the charge environment is highly stable in the dark. We note that the thermal energy at our cryostat operation temperature is $kT = 0.31$ meV, which can largely suppress thermal-induced detrapping processes unless the trap level is extremely close to the band edges. 
% Earlier results have made a similar observation, where spectral diffusion is minimal up to 3 ms dark waiting time \cite{zhang2025laser}. 
Despite the stability in the dark, the resonant excitation for PLE spectroscopy can induce charge fluctuations and cause linewidth broadening \cite{zhang2025laser, bowness2025laser}. To demonstrate this effect, we align the resonant laser pulse right after the above-band pulse and sweep the pulse width $t_\text{PLE}$. Longer resonant laser pulse diminishes the linewidth narrowing [Fig.~\ref{fig3:TimeDynamic}(e)] and center shift [Fig.~\ref{fig3:TimeDynamic}(f)], with time constants of $4.3\pm0.4\,\mu$s and $6.6\pm0.4\,\mu$s, respectively. These time constants decrease with the resonant laser power (Appendix~\ref{appendix:timeConstantVsPower}). 
We note that the linewidth change dynamics time constant is about $\sim0.7$ times that of the center shift in [Fig.~\ref{fig3:TimeDynamic}(e)(f)], which shows the opposite trend to the [Fig.~\ref{fig3:TimeDynamic}(a)(b)] that has a factor of $\sim1.5$. The different time constant ratios are explained in our modeling (Appendix \ref{appendix:modeling_details}).
\section{Numerical modeling of linewidth narrowing effect}\label{section:modeling}

\begin{figure}[t]
    \centering
    \includegraphics[scale=1]{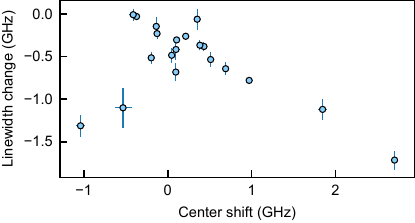}
    \caption{
        \textbf{ Survey of T center response under above-band excitation.}
        A survey of~the maximum frequency shift and linewidth change for 20 bus-waveguide-coupled single T centers, 
        % all located in the bus waveguide region of the device, 
        under above-band excitation ($\lambda_\text{AB} = 980$ nm, $t_\text{AB} = 1000$ ns, $t_\text{PLE} = 500$ ns, $t_\text{d} = 0$ ns).
        }
    \label{fig4:Survey}
\end{figure}

Additionally, we investigate similar linewidth narrowing effect beyond Itokawa. The survey includes 20 bus-waveguide-coupled T centers on the same sample [Fig.~\ref{fig4:Survey}]. The absolute linewidth narrowing and center frequency shift are correlated, which suggests that they likely originate from the same physical process. Moreover, in the large center shift regime, they show linear dependence with each other. We note that the linewidth narrowing also weakly depends on the initial linewidth at $P_\text{AB} = 0$ (Appendix~\ref{appendix:statistics}). Importantly, both blue and red center shifts are observed. This rules out the uniform charge noise model \cite{alizadehherfati2025electrical}, which only predicts blue shifts due to the quadratic Stark shift of the T center \cite{clear2024optical, alaerts2025first}. 
% which is consistent with the charge stabilization scheme mentioned earlier. 
% However, the effect shows no strong dependence on the initial linewidth (e.g., at $P_\text{AB} = 0$) and T center spectral position (Appendix~\ref{appendix:trap_survey}). 

Lastly, we construct a numerical model to explain the observed linewidth narrowing and center shift under above-band optical excitation. We attribute the effect to the laser-modulated active charge trap density $n(t)$ change. The above-band excitation generates free carriers, filling and neutralizing charge traps. The resulting mean field change induces a center shift, while the suppressed field fluctuations lead to linewidth narrowing of the T center. We assume each T center is surrounded by $N_\text{t}$ active charge traps (with $n(t)$ filled given a laser excitation configuration), which are randomly distributed spatially.
% and form a \textit{nearly} uniform spatial distribution. 
We build a rate equation for the filled trap population $p(t) = n(t)/N_\text{t}$ as,
\begin{equation}
    \label{eq:rate_equation}
        \frac{dp}{dt} = (\eta^\text{fill}_\text{AB}P_\text{AB}^\text{eff} + \Gamma^\text{fill}_\text{PLE})(1 - p)- (\Gamma^\text{detrap}_\text{PLE} + \Gamma^\text{detrap}_\text{dark})p.
\end{equation}
We assume the above-band laser has a net filling rate (i.e., filling $\gg$ detrapping) that is linearly dependent on the effective above-band laser power as $\eta^\text{fill}_\text{AB}P_\text{AB}^\text{eff}$, where $P_\text{AB}^\text{eff}$ is Auger-corrected effective power (Appendix \ref{appendix:modeling_details}). Similarly, $\Gamma^\text{fill}_\text{PLE}$ and $\Gamma^\text{detrap}_\text{PLE}$ correspond to the filling and detrapping rates related to the resonant laser, respectively. We keep the resonant laser power constant in the experiments, thus dropping the power dependence of $\Gamma^\text{fill}_\text{PLE}$ and $\Gamma^\text{detrap}_\text{PLE}$. $\Gamma^\text{detrap}_\text{dark}$ is the thermal detrapping rate. We introduce the variable $p_0$ to represent the steady-state trap filling with only the resonant laser for measuring the PLE spectrum.
% The details of the modeling process are written in Appendix \ref{appendix:modeling_details}.  

The T center ZPL response to electric field is described via the DC Stark effect \cite{clear2024optical, alaerts2025first},
\begin{equation}
    \nu(\boldsymbol{E}) = \nu_0 + \mu \boldsymbol{E} + \frac{1}{2} \boldsymbol{E} \alpha  \boldsymbol{E}^T, 
    \label{eq:Stark}
\end{equation}
% We are using results from 'First-Principles': mu,alpha should have oppossing sign
where $\alpha = -12.7\,\text{GHz/(V/$\mu$m)$^2$)}$ and $\mu \approx 0.79\,\text{D} = 4\,\text{GHz/(V/$\mu$m))}$ are the change of the polarizability tensor and dipole moment, respectively \cite{alaerts2025first}. We take the major components of $\alpha$ and $\mu$, and ignore their anisotropy to simplify the modeling process. The constant $\nu_0$ is the T center frequency without any Stark shift. We write down the electric field the T center experiences as $\boldsymbol{E} = \boldsymbol{E}_\text{DC} + \Sigma_i \boldsymbol{e}_i X_i (n,t)$, where $\boldsymbol{E}_\text{DC}$ is the constant built-in electric field due to e.g., surface charges in native oxide \cite{sun1980electron, sharma2015characterizing}
% \cite{pirro2016interface, o2001si}} 
and $\Sigma_i \boldsymbol{e}_i X_i (n,t)$ is the vector field sum of all trap generated field ($\boldsymbol{e}_i$) with $X_i (n,t) = 0 \,\text{ or } 1$ representing whether the trap is filled (neutralized) or empty (active), respectively. We define the zero-mean residue trap field noise at each trap filling point as $\boldsymbol{E}_\text{trap}(p, t) = \Sigma_i  \boldsymbol{e}_i(X_i (p,t)-\langle X_i (p,t) \rangle)$, where $\langle X_i (p,t) \rangle = 1-p$. The total electric field can then be written as,
\begin{equation}
    \boldsymbol{E} = \boldsymbol{E}_0 + p\boldsymbol{F} + \boldsymbol{E}_\text{trap}(p, t),
    \label{eq:totalE}
\end{equation}
where $\boldsymbol{E}_0 = \boldsymbol{E}_\text{DC} - \boldsymbol{F}$ and $\boldsymbol{F} = -\Sigma_i \boldsymbol{e}_i$. We note that $\boldsymbol{F}$ is a constant that changes for different T centers, while $\boldsymbol{E}_\text{DC}$ varies 
% for different T centers 
depending on their spatial locations in the bus waveguide. The center shift can be derived as,
\begin{equation}
    \Delta\nu(p) \approx (\mu+\alpha \boldsymbol{E}_0)\boldsymbol{F} (p-p_0) = C_\text{f}(p-p_0).
    \label{eq:centerShift}
\end{equation}
Under the same formalism, the change of the linewidth can be obtained by,
\begin{equation}
    \Delta \sigma(p) \approx 2\sqrt{2\ln2}|(\mu + \alpha \boldsymbol{E}_0)|(\omega(p) - \omega(p_0)),
    \label{eq:linewidthChange}
\end{equation}
where $\omega(p)$ is the standard deviation of the fluctuating $\boldsymbol{E}_\text{trap}(p, t)$. Detailed mathematical derivations can be found in Appendix~\ref{appendix:modeling_details}. In our time-resolved PLE spectroscopy, the typical repetition of the elementary pulse sequence is $N = 2\times10^5$ [Fig.~\ref{fig2:PL_response}(a)]. After pulsed laser excitation, the detector is gated on for the ``dark'' fluorescence collection window of 4.8 $\mu$s, during which the charge traps stay nearly stable [Fig.~\ref{fig3:TimeDynamic}(c)(d)]. Essentially, the T center experiences the stroboscopically sampled spectral diffusion with $N$ different charge trap configurations. For a specific T center with trap filling $p$, these configurations share the same average number of unfilled traps $(1-p)N_\text{t}$ but different spatial distributions. We approximate the trap related fluctuating fields in such a stroboscopic process by the Holtsmark distribution \cite{holtsmark1919verbreiterung, chandrasekhar1943stochastic}. In the Holtsmark picture (Appendix~\ref{appendix:TrapSpacing}), under the influence of randomly distributed charges in 3D with different spatial geometries, the FWHM linewidth ($\sigma$) of an atomic optical transition with a linear Stark shift depends on the charge density ($\rho$) as $\sigma \propto \rho^{2/3}$. 
% where $c$ is the linear Stark coefficient, $\rho$ is the charge density, and $\epsilon = e/(4\pi\epsilon_0\epsilon_r)$
Under this approximation, we can rewrite Eq.~\ref{eq:linewidthChange} as,
\begin{equation}
    \Delta \sigma(p) = C_\text{L}[(1-p)^{2/3} - (1-p_0)^{2/3}].
    \label{eq:linewidthChange_withHoltsmark}
\end{equation}
Both terms $C_\text{f}$ and $C_\text{L}$ are treated as fitting parameters.  
% \red{We note that the extra exponent for the linewidth dependence on trap filling $p(t)$ matches with the observed $\times1.5$ slower dynamics of linewidth narrowing compared to the center shift, given that $p(t)$ follows the form of $1-e^{-t/\tau}$ [Fig.~\ref{fig3:TimeDynamic}(a)(b)]. A detailed discussion can be found in Appendix \ref{appendix:modeling_details}}

Based on the established rate equation for trap filling (Eq.~\ref{eq:rate_equation}) and how the center shift (Eq.~\ref{eq:centerShift}) and linewidth narrowing (Eq.~\ref{eq:linewidthChange_withHoltsmark}) depend on it, we perform a global fitting of all Itokawa measurement data, including [Fig.~\ref{fig2:PL_response}(c)(d)] and [Fig.~\ref{fig3:TimeDynamic}]. The model perfectly fits the measurement results as shown in [Fig.~\ref{fig:sup_model_result}] (Appendix~\ref{appendix:modeling_details}), with the fitting parameters listed in Table.~\ref{table:fit_results}. % ``the descending wings''
The model not only successfully reproduces all the Itokawa's measurement data under a global fitting, it also elucidates features shown in the survey plots [Fig.~\ref{fig4:Survey}; Fig.~\ref{fig:sup_statisitcs} in Appendix~\ref{appendix:statistics}]. At far-saturated $P_\text{AB}$ powers ($p\sim1$), the saturated center shift $\Delta\nu_\text{sat} \approx (\mu+\alpha \boldsymbol{E}_0)\boldsymbol{F}$ and the linewidth change $\Delta \sigma_\text{sat} \approx -2\sqrt{2\ln2}|\mu + \alpha \boldsymbol{E}_0|\omega(p_0)$. 
T centers at different physical locations inside the waveguide experience varying $\boldsymbol{E}_0$, which is uncorrelated to the bulk trap-related field $\boldsymbol{F}$, enabling bipolar center shifts for different T centers. 
The defect-to-defect varying~$|\mu + \alpha \boldsymbol{E}_0|$ weakens the otherwise perfectly linear relation between $\Delta\sigma_\text{sat}$ and initial linewidth $\sigma_0 = 2\sqrt{2\ln2}\omega(p_0)$ (without above-band excitation) [Fig.~\ref{fig:sup_statisitcs}]. 
We can write $\Delta\sigma_\text{sat} = \pm (\sigma_0/|\boldsymbol{F}|)\Delta\nu_\text{sat}$, where $\sigma_0/|\boldsymbol{F}|$ stays close to a constant for different T centers with similar surrounding charge densities since they are correlated quantities governed by the Holtsmark distribution. This relation is manifested as the linear ``descending wings'' shown in [Fig.~\ref{fig4:Survey}]. 
% \red{cloud description}
% The apex of the survey scatter plot deviates 

\setlength{\tabcolsep}{8pt} % default = 6pt
\renewcommand{\arraystretch}{1.25} % default = 1, stretch row spacing 
\begin{table}
    \caption{
    \textbf{Global fitting results for Itokawa.}
    The uncertainty range for each variable represents the standard-error of the fitting process. 
    }
    \begin{ruledtabular}
    \begin{tabular}{ccc}
    % \hline
    \textbf{parameter} & \textbf{value} & \textbf{unit}\\
    % \hline 
    \colrule
       $\eta^\text{fill}_\text{AB}$ & $(1.33 \pm 0.08) \times10^{-4}$   & ns$^{-1}\mu$W$^{-1}$ \\
       $\Gamma^\text{fill}_\text{PLE}$ & $(3.53 \pm 1.59) \times 10^{-6}$ & ns$^{-1}$ \\ 
       $\Gamma^\text{detrap}_\text{PLE}$ &  $(1.24 \pm 0.06) \times 10^{-4}$ & ns$^{-1}$ \\ 
       $\Gamma^\text{detrap}_\text{dark}$ & $(2.10 \pm 0.57) \times 10^{-7}$ & ns$^{-1}$ \\
       $C_\text{f}$ & $0.984 \pm 0.008$ & GHz \\ 
       $C_\text{L}$ & $0.847 \pm 0.014$ & GHz \\ 
       $c_\text{aug}$ & $0.0068 \pm 0.0013$ & $\mu$W$^{-2}$ \\
    % \hline
    \end{tabular}
    \end{ruledtabular}    
    \label{table:fit_results}
\end{table}

\section{Discussion}

% We now discuss the measurement results and numerical modeling. 
% ``why not reaching lifetime limited linewidth''
From the survey, the minimum optical linewidth reaches $\sigma^\text{min}_\text{sat} \sim 360$ MHz under above-band excitation, which is larger than the thermal broadening at $T = 3.6$ K ($\sigma_\text{th} \sim$ 120 MHz) \cite{bergeron2020silicon, johnston2024cavity}. Under a far-saturated power $P_\text{AB} = 53.1 \,\mu$W, the steady-state trap occupancy when above-band and resonant laser pulses perfectly overlapped is $p_\text{ss} \approx 0.954$ based on the fitted parameters in Table.~\ref{table:fit_results}. The $\sigma_\text{sat}$ is close to the homogeneous linewidth measured via spectral hole burning for device-coupled single T centers at $T = 4$ K \cite{zhang2025laser}.
% Considering the near-unity trap filling and $\Gamma_\text{sat} > \Gamma_\text{th}$, we speculate that there exist fluctuating charge traps that cannot be suppressed by above-band excitation at $\lambda_\text{AB} = 980$ nm. 
We note that the average heating power for the above-band laser in the pulsed experiment is $P_\text{AB}^\text{avg} = 3.1\,\mu$W (surface reflection ignored). From a separate cryostat MW-induced heating test, we extract a platform cooling power of 8.5 $\mu$W/mK. 
% Even we assume all above-band laser power is absorbed by the silicon substrate (the device layer absorption is $\sim$ 10$^{-4}$), 
The amount of laser-induced average heating is negligible. The device layer can absorb a small fraction ($\sim 10^{-4}$) of the above-band laser, we estimate a peak temperature increase of $\sim 0.14$ K based on a COMSOL model. This transient heating is too small to account for the difference between $\sigma_\text{sat}^\text{min}$ and $\sigma_\text{th}$ at $T=3.6$ K.   When only resonant excitation is present, the trap filling steady-state reaches $p_0 \approx 0.028$, leaving a large number of unfilled traps contributing to the linewidth. In the dark, the detrapping time $\tau_\text{dark} = 1/\Gamma^\text{detrap}_\text{dark} = 4.8 \pm 1.3 $~ms, which is consistent with the reported value \cite{zhang2025laser}.

% ``assumption and limitation of the model''
% ``why rule out surface traps
We now discuss the assumptions and limitations of the model. As mentioned earlier, we approximate the stroboscopic sampling of different charge configurations by the Holtsmark distribution. The ideal Holtsmark physics deals with a large number of randomly distributed, and moving charges, such that the field average over time gives $\langle \boldsymbol{F} \rangle= 0$.  
In our bus waveguide structure, we estimate the bulk trap density to be $\sim 8\times10^{15}$ cm$^{-3}$ based on the typical initial linewidth $\sigma_0 = 1.5$ GHz (Appendix~\ref{appendix:TrapSpacing}), which corresponds to 85 traps in a 220 nm wide cube. 
Although our bulk traps have random spatial positions, low trap density and the finite nature of our waveguide geometry enables $\boldsymbol{F} \neq 0$. 
This assertion is bolstered by the fact that due to the dominant quadratic Stark effect, if $\boldsymbol{F} = 0$, only a blue shift can be observed (Appendix~\ref{appendix:modeling_details}), which contradicts our measurements [Fig.~\ref{fig4:Survey}]. The model has ignored the anisotropic Stark effect of the T center \cite{clear2024optical, alaerts2025first} for simplicity. When included, the anisotropic $\mu$ and $\alpha$ will complicate the Stark response and the calculation of $|\mu + \alpha \boldsymbol{E}_0|$. Some of the T centers we observe have near-zero linewidth narrowing but non-zero center shift (or vice versa) [Fig.~\ref{fig4:Survey}]. Our current model cannot well explain these cases, which may necessitate the full consideration of the anisotropy.
% \red{The perfect Holtsmark model deals with a large number of randomly distributed charges with $\boldsymbol{F} = \Sigma_i \boldsymbol{e}_i = 0$. In our bus waveguide structure, we estimate the bulk trap density of $3\times10^{15}$ cm$^{-3}$ based on the typical initial linewidth $\sigma_0 = 1.5$ GHz, which corresponds to 32 traps in a 220 nm wide cube. Even though the traps are generated in random spatial positions, the finiteness and deviation from a perfectly symmetric distribution enables $\boldsymbol{F} \neq 0$. We note that if $\boldsymbol{F} = 0$, due to the quadratic Stark effect, only blue shift can be observed (Appendix:~\ref{appendix:modeling_details}), which contradicts our measurements [Fig.~\ref{fig4:Survey}].} 

The Holtsmark exponent for the charge density scales with the dimensionality ($d$) of the charge distribution as $2/d$ \cite{samama2023statistics}. If the major contribution of the linewidth narrowing of Itokawa is caused by the suppression of surface charge traps ($d=2$), the linewidth narrowing and center shift dynamics would always share the same time constant, which deviates from the experimental observation [Fig.~\ref{fig3:TimeDynamic}]. However, for T centers that are very close to the surfaces, the contribution from the surface charge traps can be dominant, which the current model cannot describe. We note that our measurement when probing T centers may introduce a selection bias as the waveguide polarization will well couple to only a subset of T center orientations, and we typically skip T centers that have a very broad initial linewidth. 

Another potential pathway is to model the $N_\text{t}$ traps as identical Bernoulli variables \cite{alizadehherfati2025electrical}, where the linewidth primarily scales with $\sqrt{p(1-p)}$. All the measured T centers that showed narrowing effect undergo monotonic linewidth reduction with above-band excitation, indicating the initial trap occupancy $p_0\geq \sim 0.5$ within this model. However, even with this constraint, we note that such a model fails to describe Itokawa's data under a global fitting. One potential deviation of this model from the reality is that above-band excitation at $\lambda_\text{AB} = 980$ nm fills traps in different ways due to e.g., the inhomogeneity of the trap energy levels \cite{troxell1983ion}.

\section{Outlook and conclusions}

We now discuss pathways to further narrow the T center linewidth. 
% The saturated optical linewidth (Fig.\ref{fig2:PL_response}c) for device-coupled single T centers is limited by thermal broadening at $T = 3.6$ K in our cryostat and remaining charge fluctuations due to unfilled traps and the resonant laser excitation \cite{zhang2025laser, bowness2025laser}. 
Beyond minimizing thermal broadening at a lower temperature \cite{bergeron2020silicon}, surface passivation \cite{aberle2000surface} and more gentle ways of T center formation can be utilized to decrease the surface and bulk trap densities. 
% Built upon the narrowing scheme we showed in this work, further linewidth reduction can be obtained when operating at a lower temperature to eliminate thermal broadening \cite{bergeron2020silicon} and in an engineered silicon host with lower surface and bulk trap densities. For passivating the surface charge traps, such as silicon dangling bonds, hydrofluoric acid wet etch (to form Si-H bonds) and thin layer ($<$ 10 nm) Al$_2$O$_3$ atomic layer deposition \cite{dingemans2012status} can be utilized. 
T center generation can be realized without the violent ion implantation method. By leveraging native carbon and hydrogen atoms in the silicon, T centers can be formed using electron irradiation \cite{safonov1996interstitial, bergeron2020silicon} and only thermal annealing \cite{minaev1981thermally, lightowlers1994hydrogen} methods. These more gentle methods will likely lower the charge trap density by decreasing formation of point defects (vacancies and interstitials) and defect complexes, which are typically present during ion implantation processes. From our modeling, another pathway to decrease the linewidth is to minimize $|\mu + \alpha \boldsymbol{E}_0|$. Due to the respective 2D and 3D nature of $\boldsymbol{E}_0$ for waveguide and nanobeam PC cavity structures, external vector DC electric field control is required to enable $|\mu + \alpha \boldsymbol{E}_0| = 0$ for waveguide- or cavity-coupled single T centers.

In summary, we have demonstrated linewidth narrowing for device-coupled single T centers by using above-band optical excitation to suppress environmental charge fluctuations. The work tackles an important roadblock for the T center platform toward quantum networking applications. We expect that a sub-100 MHz linewidth is within reach in the near term, which is close to the Purcell-enhanced radiative linewidth. This will enable immediate applications for high-fidelity single-shot spin readout \cite{wong2025cavity} and fast indistinguishable telecom photon generation \cite{ourari2023indistinguishable}. 
% The narrower linewidth can lead to larger atom-cavity cooperativity. With realistic system parameters such as cavity $Q = 2\times10^5$, we estimate the cooperativity $C > 1.2$ for T center linewidth $\Gamma < 100$ MHz.
Moreover, the narrower linewidth leads to a higher cooperativity for cavity-coupled single T centers, enabling crucial networking protocols including spin-photon entanglement using time-bin photonic qubits \cite{nguyen2019integrated}, remote spin entanglement via a sender-receiver network topology \cite{knaut2024entanglement, beukers2024remote}, and hybrid quantum networking \cite{chai2026direct}.

\begin{acknowledgments}
We gratefully acknowledge Yizhi Zhu and Geoffroy Hautier for helpful discussions. Support for this research was provided by the National Science Foundation (NSF) CAREER Award (No. 2238298) and Electronics, Photonics and Magnetic Devices (EPMD) program (No. 2527905), as well as the Robert A. Welch Foundation (Grant No. C-2134). A.J. acknowledges support from the NSF NRT BRIDGE-CQED program (No. 2346014). We acknowledge the use of cleanroom facilities supported by the Shared Equipment Authority at Rice University.
\end{acknowledgments}

\vspace{12pt}
\begin{center}{
\textbf{\small{DATA AVAILABILITY}}
}
\end{center}

The datasets generated and/or analyzed during the current study are available from the corresponding author on reasonable request.

\appendix
% \red{[SC] Let's arrange the appendix sections based on the order they are referred in the main text.}

\section{CAVITY-ENHANCED RADIATIVE LINEWIDTH}\label{appendix:lifetimeLimitedLinewidth}

\begin{figure}[H]
        \centering
        \includegraphics[scale=1]{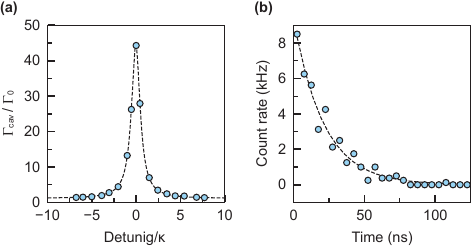}
        \caption{\textbf{Purcell effect for T center ``Earth''.} 
            \textbf{(a)} Decay rate enhancement at different cavity detunings. The black dashed line shows the Lorentzian fitting with a linewidth of 1.19 $\pm$ 0.01 $\kappa$. The slightly larger fitted linewidth than the cavity linewidth ($\kappa$) is due to the dephasing and spectral diffusion of the T center \cite{johnston2024cavity}.
            % kappa = 3.82 GHz
            \textbf{(b)} The shortest enhanced fluorescence lifetime observed for the T center. A single exponential fitting reveals a lifetime of $21.2 \pm 0.21$ ns.
            }
        \label{figSI:supH_earthPurcell}
    \end{figure}

Utilizing the cavity quantum electrodynamics, Purcell-enhanced radiative linewidth ($\Gamma_\text{cav}$) can exceed the intrinsic lifetime-limited linewidth ($\Gamma_0/2\pi=$ 0.17~MHz) for a cavity-coupled single T center. The low-loss and small mode-volume optical cavity, and perfect dipole alignment with the cavity field polarization promote a stronger Purcell effect. The highest Purcell factor we have ever observed for a single T center (named ``Earth'') is $F_p = 43.3$ [Fig.~\ref{figSI:supH_earthPurcell}(a)], corresponding to a fluorescence lifetime of $\tau_\text{cav} = 21.2$ ns [Fig.~\ref{figSI:supH_earthPurcell}(b)] and an enhanced radiative linewidth of $\Gamma_\text{cav}/2\pi = 7.5$ MHz. The estimated atom-cavity coupling strength for this T center is $g/2\pi = 88.6$ MHz.  

% The intrinsic lifetime limited linewidth is 169 kHz, derived from the intrinsic optical lifetime of 940 ns. The 
% The largest Purcell effect we have ever observed for a single T center is around a factor of 44.3, corresponding to optical lifetime of 21.2 ns and lifetime limited linewidth of 7.5 MHz, as shown in Fig.\ref{figSI:supH_earthPurcell}. 

\begin{figure*}[t]
    \centering
    \includegraphics[scale=1]{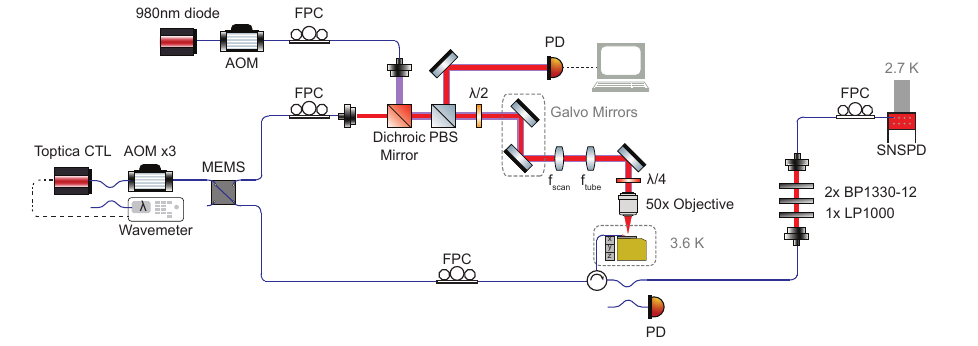}
    \caption{\textbf{Experimental setup details.} AOM: acousto-optic modulator. BP: bandpass filter. LP: longpass filter. FPC: fiber polarization controller.  MEMS: micro-electro-mechanical system. PBS: polarizing beam splitter. PD: photodiode. SNSPD: superconducting nanowire single photon detector. 
    }
    \label{fig5:detailed_setup}
\end{figure*}

    % \onecolumngrid

\section{EXPERIMENTAL SETUP}\label{appendix:detailedSetup}

In this section we provide a more detailed description of our experimental setup, shown in [Fig.~\ref{fig5:detailed_setup}]. The frequency stabilized, fiber-coupled telecom CW tunable laser (Toptica CTL 1320) is sent through three AOMs to generate excitation pulses, controlled by arbitrary waveform generator (Zurich HDAWG 750 MHz). Another 980 nm laser diode (Thorlabs LP980-SF15) is also fiber coupled and pulsed via an AOM (AeroDIODE 940AOM-1). We control the optical power of the former with a series of neutral density (ND) filters, and the latter by modulating the AOM driving power.
We combine both lasers using a dichroic mirror (Thorlabs DMLP1000, 1000 nm cut-on wavelength). 
A polarizing beam splitter (PBS, Thorlabs LPIREA050-C) and a half-wave plate (HWP, Thorlabs, WPH10M-1310) are used to prepare the desired linear polarization. 
After that, the beams pass through a 2D galvo mirror system (Thorlabs GVSM002), scan lens (Thorlabs LSM54-1310, $f$ = 54 mm), tube lens (Thorlabs TTL200MP2, $f$ = 200 mm), and 50$\times$ objective (Mitutoyo MY50X-825) before entering the top window of the cryostat and reaching the sample surface. We add a quarter wave plate (QWP, Thorlabs WPQ10M-1310) before the objective to enable reflection signal for the PD during optical alignment. Another optical path for the resonant laser to reach the nanophotonic devices is through an angle-polished fiber coupled to the GC. We can switch between the two paths using a MEMS switch (Agiltron FFSM-226C01333). The T-center fluorescence emission signal is collected via the fiber path into the SNSPD (Single Quantum EOS 210CS). We stack two bandpass (Thorlabs FBH1330-12) and one longpass (Thorlabs FELH1000) spectral filters to reject both above-band laser photons and the background fluorescence emission from the sample. 
% It is then routed via the circulator and sent into the SNSPD (Single Quantum). A previous paper from our group includes more details of our experimental setup \cite{johnston2024cavity}.

% \section{SAMPLE PREPARATION \& T CENTER GENERATION} \label{appendix:TGeneration}
% We fabricate our nanophotonic devices on a sample from the same wafer as our previous paper \cite{johnston2024cavity}. To generate T centers, we perform ion implantation (Coherent) with a fluence of 7$\times 10^{12}$ and 3.5$\times 10^{12}$ ion/cm$^2$ for $^{12}$C and $^{1}$H respectively. Between the $^{12}$C and $^{1}$H implantations, the sample undergoes a 20 s annealing at 1000 $^{\circ}$C (argon background). After the $^{1}$H implantation, the sample is placed in boiling water for 1 hour, then annealed at 460 $^{\circ}$C for 180 s in a nitrogen background. 

\section{BUS WAVEGUIDE DEVICE DESIGN}\label{appendix:meanderDesign}
\begin{figure}[b]
    \centering
    \includegraphics[scale=1]{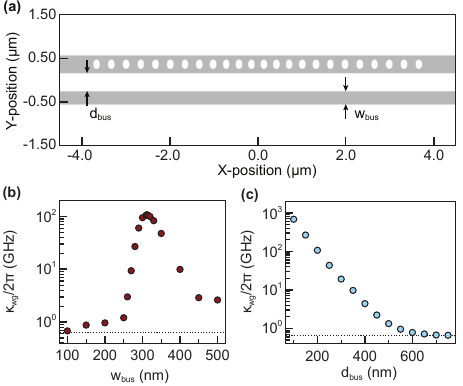}
    \caption{
    \textbf{Bus waveguide design.}
    \textbf{(a)} Schematic of the bus waveguide structure. 
    % with spacing ($d_\text{bus}$) and bus waveguide width ($w_\text{bus}$). 
    \textbf{(b)} Simulated loss rates of the bus-waveguide-coupled PC cavity with varying $w_\text{bus}$ while keeping $d_\text{bus} = 200$ nm. $\kappa_\text{wg}$ is maximum at $w_\text{bus} = 313$ nm.
    \textbf{(c)} Similar as panel \textbf{(b)}, but sweeping $d_\text{bus}$ while keeping $w_\text{bus} = 313$ nm.
    The simulation geometry uses $N_\text{cav} = 12$ holes and $N_\text{mir} = 6$ holes on each side of the cavity.
    The dashed gray line shows the loss rate from an isolated PC cavity with the same parameters.
    }
    \label{fig:sup7_buswg}
\end{figure}
A bus waveguide device layout [Fig.~\ref{fig:sup7_buswg}(a)] is implemented for this experiment, allowing multiple PC cavities to couple to a single GC. This greatly improves the fabrication yield and the likelihood of observing cavity-coupled T centers. We use the same 1D PC cavity design as shown in our earlier work \cite{johnston2024cavity}.
% parameters $a_\text{mir}, \text{ }a_\text{cav},\text{ }w_\text{PC}  = (334, 282, 410)$ nm to define the PC cavities \cite{johnston2024cavity}. 
% In the fabricated devices, a parabolic taper with length $N_\text{cav} = 12$ holes is used to transition between the band gap ($a_\text{mir})$ and band-edge ($a_\text{cav})$ periodicity, this shape is chosen to create an approximate Gaussian mode profile in the along the length of the PC beam \cite{tanaka2008design, johnston2024cavity}. A 'mirror region' of size $N_\text{mir} = 8$ holes separates neighboring PC cavities.
% For the fabricated device parameters, we calculate a mode volume $V_\text{mode} = 0.0287$ $\mu\text{m}^3$ or 0.159 $(\lambda/n_\text{eff})^3$. 
The PC cavity is evanescently coupled to the bus waveguide. The coupling rate between the PC cavity and bus waveguide mode is maximized when the two modes have the same effective index ($n_\text{eff}$). 
To tune $n_\text{eff}$ of the bus waveguide, we sweep its width $w_\text{bus}$. The mode matching condition is manifested as the evanescently coupled cavity reaches maximum loss rate ($\kappa_\text{wg}$) [Fig.~\ref{fig:sup7_buswg}(b)] given a constant cavity-waveguide spacing $d_\text{bus}$. By controlling $d_\text{bus}$, we can modulate $\kappa_\text{wg}$ to match the cavity loss rate ($\kappa_\text{sc}$) resulted from surface roughness and fabrication imperfections. This $\kappa_\text{wg}$ tuning capability enables us to obtain critically-coupled ($\kappa_\text{wg} = \kappa_\text{sc}$) cavities for most efficient photon extraction in the Purcell regime. The fabricated devices used in this work have $w_\text{bus} = 313$ nm and $d_\text{bus} = 450$ nm.
% The simulated loss rate is similar that of the ideal structure, but in reality imperfections such as surface roughness will cause scattering loss in fabricated devices. We refer to these loss channels due to imperfections as $\kappa_\text{sc}$, and define a parameter $\eta_\text{cav} = \kappa_\text{wg}/(\kappa_\text{wg} + \kappa_\text{sc})$ to quantify the proportion of the cavity loss that couples to the waveguide mode via $\kappa_\text{wg}$. 
% To achieve a high $\eta_\text{cav}$ we control the distance between the PC cavity and the bus waveguide $(d_\text{bus})$, the fabricated devices have $d_\text{bus} = 450$ nm. 
The bus waveguide is terminated with a Bragg reflector, creating a standing wave in the waveguide. Therefore $\kappa_\text{wg}$ is dependent not only on the mode matching between the PC and bus waveguide fields, but also the phase difference between them \cite{xu2000scattering}.

\section{PERFORMANCE OF LASER SCANNING MICROSCOPE}\label{appendix:LSMSpotSize}

In the experiment, we use a laser scanning microscope (LSM) to locate and excite device-coupled T centers. 
% Therefore, we need to evaluate its performance by measuring its resolution. 
The theoretical limit of the lateral and axial resolution for a conventional microscope is defined as \cite{Wilhelm2024Zeiss}:
\begin{equation}
\begin{aligned}
     \Gamma_{x,\,y} &= \frac{0.51 \lambda}{\text{NA}},&&
     \Gamma_{z} = \frac{1.67n\lambda}{\text{NA}^2},
\end{aligned}
\end{equation}
where $\lambda$ is the laser wavelength, NA is the numerical aperture of the objective, and $n=1$ is the refractive index of medium, which is vacuum in this case. For our system, the theoretical resolution for the resonant ($\lambda_\text{PLE} = 1326$ nm) and above-band ($\lambda_\text{AB} = 980$ nm) wavelengths are $\Gamma_{x,\,y}^{\text{PLE}} = 1.6$ $\mu$m, 
$\Gamma_{z}^{\text{PLE}} = 12.6$ $\mu$m
and
$\Gamma_{x,\,y}^{\text{AB}} = 1.2$ $\mu$m,  $\Gamma_{z}^{\text{AB}} =~9.3$~$\mu$m, respectively.

% Next, we experimentally measure the axial and lateral resolution.
% We rely on a gold cross that is 10 $\mu m$ wide (significantly larger than resolution), and is 220 nm higher than the silicon sample surface. 
To characterize the lateral resolution of the microscope, we utilize a 10 $\mu$m wide gold cross (thickness 220~nm) on the silicon surface.
When a Gaussian beam hits the interface from a normal angle, the reflected power near the silicon-gold interface, can be derived as:

\begin{equation}
\begin{aligned}
    \label{PSF derivation}
    P_{\text{rise}} (x') &= R_\text{Au}I_0 \int_{x_0-x'}^{\infty} e^{\left(-2x^2/w^2\right)}dx \int_{-\infty}^{\infty} e^{\left(-2y^2/w^2\right)}dy
    \\
    &+ R_\text{Si}I_0 \int_{-\infty}^{x_0-x'} e^{\left(-2x^2/w^2\right)}dx \int_{-\infty}^{\infty} e^{\left(-2y^2/w^2\right)}dy
    \\
    &= \frac{I_0\pi w^2}{4}(R_\text{Si}+R_\text{Au})
    \\
    &+\frac{I_0\pi w^2}{4}(R_\text{Si}-R_\text{Au}) \operatorname{erf}\left(\sqrt{2}\frac{x_0-x'}{w}\right)
\end{aligned}
\end{equation}
\noindent where $w$ is the $1/e^2$ radius, $R_\text{Si}$ and $R_\text{Au}$ are the reflectivity of silicon and gold at the laser wavelength, $x_0$ and $x'$ are the positions of the gold rising edge and center of the laser beam, respectively, and $I_0$ is the peak intensity of a Gaussian beam at center ($x=y=0$). 
We ignore the interference, scattering (e.g., from the corner and sidewall of the gold layer), and the geometric shadowing effect in the derivations above. The geometric shadowing causes minor ``pre-step'' dips in the measurement. Similarly, we can obtain an expression for a falling edge at $x_1$. For fitting purpose, we can absorb reflectivity and intensity into a coefficient and a constant offset. Therefore, the final position-dependent reflection can be written as,
\begin{equation}
\begin{aligned}
    P(x') &= A\left[\operatorname{erf}\left(\sqrt{2}\frac{x_0-x'}{w}\right)+\operatorname{erf}\left(\sqrt{2}\frac{x'-x_1}{w}\right)\right]+C.
\end{aligned}
\end{equation}
\begin{figure}[t]
    \centering
    \includegraphics[scale=1]{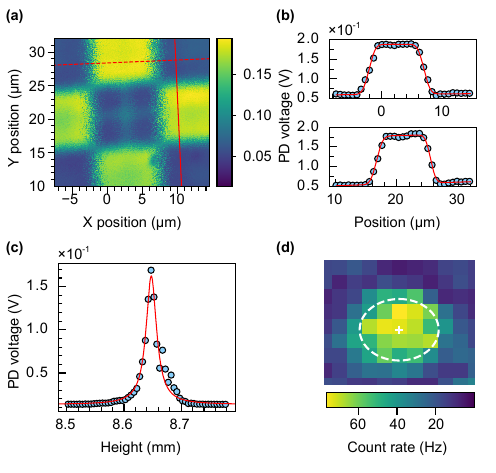}
    \caption{\textbf{LSM system performance at $\lambda_\text{PLE} = 1326$ nm.} 
    % The figures showed are all measured at 1326 nm
    \textbf{(a)} 2D reflection scan and line cuts.
    \textbf{(b)} Reflection signal along the line cuts in panel (a). The top (bottom) one is taken along rotated $x$ ($y$) axis. The fitted $\Gamma_x^{\text{PLE}}$, $\Gamma_y^{\text{PLE}}$ = 1.7 $\pm$ 0.3, 1.6 $\pm$ 0.1 $\mu$m.
    \textbf{(c)} Axial resolution scan and Lorentzian fit (red line), revealing $\Gamma_{z}^{\text{PLE}}$ = 21.8 $\pm$ 0.3 $\mu$m.
    \textbf{(d)} 2D PLE scan for a single T center fluorescence emission. 
    % collected from GC. 
    Each pixel is $0.5 \times 0.5$ $\mu$m$^2$. Gaussian fitting yields FWHM spot size $\phi_x=2.6 \pm 0.1$ $\mu$m, and $\phi_y=2.1\pm 0.1$ $\mu$m.
    % \Gamma_x^{\text{PLE}}$, $\Gamma_y^{\text{PLE}}$ = 2.6, 2.1 $\mu$m.
    }
    \label{fig6:LSM}
\end{figure}

We then fit the reflection signal when the laser is scanned across the gold cross using the above equation [Fig.~\ref{fig6:LSM}(a)(b)]. A minor rotation ($\sim 2^{\circ}$) is observed between the sample and the galvo mirror system. The FWHM diameter is obtained by $\sqrt{2\ln2}w$.
% from edges and extract $\Gamma$ using the following equations, considering experimental background signal and reflectivity of the materials at the edge:
% Since our sample has a small rotation with respect to scanning microscope system, we take multiple line cuts from rotated x and y axis to fit $\Gamma_x$ and $\Gamma_y$, and take the average as measured lateral resolution (Fig. 6a and b). 
% for the $x$ direction, and similarly for $y$ direction. The $x$ and $y$ directions are defined with respect to the sample, which is slightly rotated from the microscope [Fig. \ref{fig6:LSM}(a)(b)]. 
To fit $\Gamma_{x,\,y}$, we average over multiple line-cuts in the same axis.
We obtain lateral resolution for $\lambda_{\text{PLE}}$ = 1326 nm to be $\Gamma_{x}^{\text{PLE}}$, $\Gamma_{y}^{\text{PLE}}$ = 1.7 $\pm$ 0.3 $\mu$m, 1.6 $\pm$ 0.1~$\mu$m; for $\lambda_{\text{AB}}$ = 980 nm to be $\Gamma_{x}^{\text{AB}}$, $\Gamma_{y}^{\text{AB}}$ = 1.3 $\pm$ 0.1 $\mu$m, 1.2 $\pm$ 0.2 $\mu$m. These fitted spot sizes are close to the diffraction limit. We attribute the slight asymmetry to the beam path misalignment. We also measure axial resolution by fitting a Lorentzian profile to a 1D scan in the $z$ direction. We obtain $\Gamma_{z}^{\text{PLE}}$, $\Gamma_{z}^{\text{AB}}$ = 21.8 $\pm$ 0.3 $\mu$m, 23.8 $\pm$ 1.0 $\mu$m [Fig.~\ref{fig6:LSM}(c)]. When performing spatially-resolved PLE spectroscopy, we observed a larger asymmetric fluorescence spot than the laser spot size [Fig.~\ref{fig6:LSM}(d)], which have contributions from T center emission saturation and the asymmetric light coupling to the waveguide along different axes. 
% The measured resolution is also supported by experimental T-center emission data , note this is not equivalent to laser spot size on sample surface. 

% In our experiment, we focus at 1326 nm and deliver both wavelength together. 
In the linewidth narrowing experiment, the LSM is at focus for $\lambda_\text{PLE} = 1326$ nm and defocused for $\lambda_\text{AB} = 980$ nm by $\sim 10$ $\mu$m. We utilize the lateral and axial resolution at focus for $\lambda_\text{AB}$ to estimate its defocused spot size of $\Gamma_{x}^{\text{AB}}$, $\Gamma_{y}^{\text{AB}} = 3.8 \pm 0.6$ $\mu$m, $4.2\pm 0.2$~$\mu$m.

\section{ITOKAWA RESPONSE UNDER ABOVE-BAND EXCITATION AT $\lambda_\text{AB} = 532$ nm}\label{appendix:532nm_response}

\begin{figure}[H]
    \centering
    \includegraphics[scale=1]{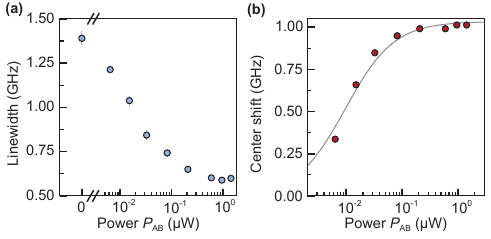}
    \caption{
        \textbf{Itokawa response under CW above-band excitation ($P_\text{AB}$) at $\lambda_\text{AB} =$ 532 nm.}
        \textbf{(a)} Gaussian-fitted FWHM linewidth. 
        The initial and minimum linewidths are 1.39 $\pm$ 0.04 GHz and 0.60 $\pm$ 0.01 GHz, respectively.
        \textbf{(b)} Center frequency shift. The frequency shift is fitted to a saturation curve (gray line) with $P_\text{sat}^{532} =$ 9.7 nW. 
        % \textbf{(c)} Gaussian-fitted amplitude, background, and integrated area (background subtracted). We observe
        % an increase in the background beginning at $\sim$0.2 $\mu$W, and 
        % a 25\% drop in peak area after $\sim$ 20 nW, suggesting above-band excitation at $\lambda_\text{AB} = 532$ nm can partially excite the T center.
        % there is minor PL fluorescence contribution to the spectra.
    }
    \label{fig:l532nm_itokawa}
\end{figure}

% \twocolumngrid
The Itokawa response is characterized under $\lambda_\text{AB} = 532$ nm [Fig.~\ref{fig:l532nm_itokawa}]. We find $P_\text{sat}^{532}$ to be two orders of magnitude lower than $P_\text{sat}^{980}$, which results from the much higher optical absorption at $\lambda_\text{AB} = 532$ nm than that at $\lambda_\text{AB} = 980$ nm at the cryogenic temperature.
% reflecting the silicon material's higher carrier generation efficiency at lower wavelengths. 
However, the maximum linewidth narrowing and center shift are consistent between $\lambda_\text{AB} = 532$ nm  and 980 nm, suggesting similar saturated trap filling in both cases.
% occupancy of the local traps is near unity in both cases. 

\section{POWER DEPENDENCE OF RESONANT LASER DYNAMICS}\label{appendix:timeConstantVsPower}

    \begin{figure}[t]
        \centering
        \includegraphics[scale=1]{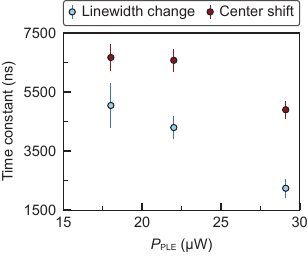}
        \caption{\textbf{Resonant laser dynamics power dependence.} 
                The time constants are derived from varying resonant laser pulse width ($t_\text{PLE}$) under different excitation powers, as performed in [Fig.\ref{fig3:TimeDynamic}(e)(f)]. 
                % There is a weak correlation in which higher excitation power corresponds to lower time constants, which is consistent with the observation discussed in Sec.\ref{subsection:linewithReduction}.
                }
        \label{figSI:supG_timeConstantVsPower}
    \end{figure}

The diminishing of the linewidth narrowing effect due to the resonant laser [Fig.~\ref{fig3:TimeDynamic}(e)(f)] is investigated under different laser excitation powers ($P_\text{PLE}$). We note that the steady-state trap filling for above-band laser only (under $P_\text{AB} \gg P^\text{sat}_\text{AB}$) and resonant laser only cases are $p\sim1$ and $p\sim0$, respectively. Using the pulse sequence for [Fig.~\ref{fig3:TimeDynamic}(e)(f)], the trap filling dynamics can be written as $p(t_\text{PLE}) \approx \exp(-t_\text{PLE}/\tau_\text{PLE})$, where $\tau_\text{PLE} = (\Gamma^\text{fill}_\text{PLE}+\Gamma^\text{detrap}_\text{PLE}+\Gamma^\text{detrap}_\text{dark})^{-1} \approx (\Gamma^\text{fill}_\text{PLE}+\Gamma^\text{detrap}_\text{PLE})^{-1}$. Assuming both the filling and detrapping due to the resonant laser excitation is linearly proportional to the power $P_\text{PLE}$, the time constant $\tau_\text{PLE} \propto 1/P_\text{PLE}$. The general trend of the observed power dependence matches with the predicted functional form [Fig.~\ref{figSI:supG_timeConstantVsPower}]. However, more measurements at lower powers are needed to confirm the predicted dependence. 
The signal-to-noise ratio prevents us from using lower powers.

\begin{figure*}[t]
    \centering
    \includegraphics[scale=1]{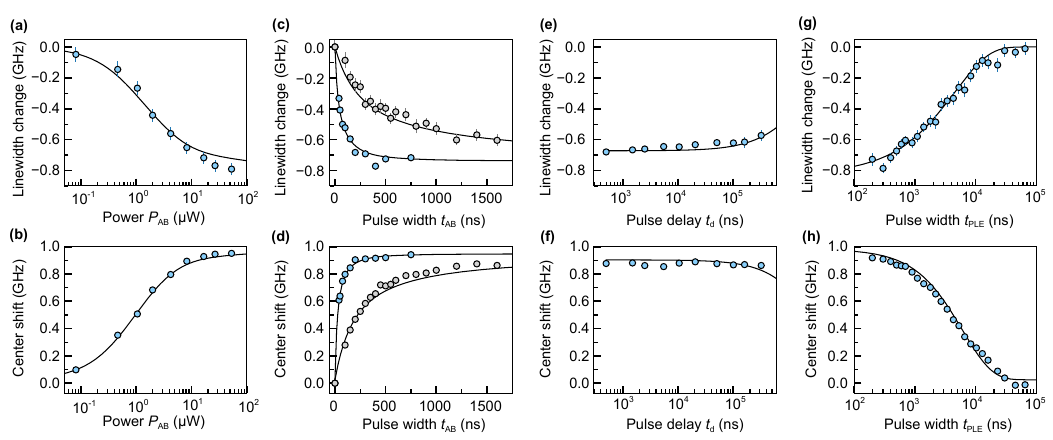}
    \caption{
    \textbf{Global fitting results.}
    The solid lines show the result of the global fitting to the rate equation model. 
    Data in panels \textbf{(a)} and \textbf{(b)} are replotted from [Fig.~\ref{fig2:PL_response}(c)(d)] and data in panels \textbf{(c)} to \textbf{(h)} are replotted from [Fig.~\ref{fig3:TimeDynamic}].
    % Panels \textbf{(a)} and \textbf{(b)} show the linewidth and center frequency dependence on $P_\text{AB}$, here $\lambda_{AB} = 980$ nm. 
    % Panels \textbf{(c)} and \textbf{(d)} show the linewidth change and frequency shift while varying $t_\text{AB}$. The blue and gray points use $P_\text{AB} = ~$2.1 and 36.1 $\mu$W respectively. 
    % Panels \textbf{(e,f)} and \textbf{(g,h)} show the dependence on $t_\text{delay}$, and $t_\text{PLE}$ respectively. 
    % A constant $P_\text{PLE}$ is used across all panels. 
    }
    \label{fig:sup_model_result}
\end{figure*}

\section{RATE EQUATION AND DETAILS OF MODELING}
\label{appendix:modeling_details}
In this section, we will provide mathematical details for the numerical modeling. The global fitting is performed based on the trap occupancy $p(t) = n(t)/N_\text{t}$ obtained from the analytical solution of Eq.~\ref{eq:rate_equation}, 
\begin{equation}
    % p = \tau (A - C~\text{exp}[-t/\tau]).
    % p(t) = A \tau + (p(0) - A\tau)\text{exp}[-t/\tau]
    p(t) = p_\text{ss} + (p(0) - p_\text{ss})\text{exp}[-t/\tau],
    \label{eq:ss_solution}
\end{equation}
where
\begin{equation}
    \begin{aligned}
        A &= \eta^\text{fill}_\text{AB}P_\text{AB}^\text{eff} + \Gamma^\text{fill}_\text{PLE},\\
        B &= \Gamma^\text{detrap}_\text{PLE} + \Gamma^\text{detrap}_\text{dark}, \\
        \tau &= 1/(A + B), \\
        p_\text{ss} &= A/(A+B).
    \end{aligned}
\end{equation}
% and $t = 0 $ signifies the start of the elementary pulse sequence. 
The above-band laser generates free carriers leading to the trap filling (Eq.~\ref{eq:rate_equation}). Regarding the free carrier generation, we consider Shockley-Read-Hall (first order) and Auger-Meitner (third order) decay pathways. The silicon PL (second-order) recombination is much weaker due to the material's indirect band gap, and it's neglected for simplicity. An ``effective power" is introduced to capture the interplay between the first and third order decay pathways, 
\begin{equation}
    \label{eq:effective_power}
    P_\text{AB}^\text{eff} = \frac{P_\text{AB}}{(1 + c_\text{aug}P_\text{AB}^2)^{1/3}},
\end{equation}
where $c_\text{aug}$ is treated as a fitting parameter.
To perform the global fit, we solve Eq.~\ref{eq:rate_equation} for each elementary pulse sequence, forming a piecewise function to describe the $p(t)$ across the pulse sequence. In order to match the experimental conditions, we repeat each pulse sequence until the final occupancy reaches a steady-state. This requires a small number of pulse repetitions ($< 100$) for the final fitted parameters (Table.~ \ref{table:fit_results}), which is much less than the $N = 2\times10^5$ repetitions used in experiment. Essentially the experimental data reflects the steady-state condition of the trap filling right after all laser excitation pulses. Figure.~\ref{fig:sup_model_result} shows the global fitting based on the model, which perfectly fit experimental results. A least-squares solver is used to carry out the parameter optimization \cite{virtanen2020scipy}. We see no dependence of the solution on our initial guess for the parameters.
% \cite{virtanen2020scipy}. 

Next, we provide details on how we derive the equations for center shift and linewidth narrowing. Substituting the $\boldsymbol{E}$ (Eq.~\ref{eq:totalE}) into the Stark shift equation (Eq.~\ref{eq:Stark}), we have the full expression for the T center frequency, 
\begin{equation}
    \begin{aligned}
        \nu(p,t) = \nu_0 &+ \mu(\boldsymbol{E}_0 + p\boldsymbol{F})+\frac{1}{2}\alpha(\boldsymbol{E}_0 + p\boldsymbol{F})^2\\
        &+[\mu+\alpha(\boldsymbol{E}_0 + p\boldsymbol{F})]\boldsymbol{E}_\text{trap}(p,t)\\
        &+\frac{1}{2}\alpha \boldsymbol{E}^2_\text{trap}(p,t).
    \end{aligned}
    % \begin{aligned}
    %     \nu(p) &= \nu_0 + \mu\boldsymbol{E}_0 + 1/2\alpha(\boldsymbol{E}_0^2 + (p\boldsymbol{F})^2 + \boldsymbol{E}_\text{trap}^2) \\
    %     &+ (\mu + \alpha\boldsymbol{E}_0)p\boldsymbol{F}
    %     + (\mu + \alpha(\boldsymbol{E}_0 + p\boldsymbol{F}))\boldsymbol{E}_\text{trap}.
    % \end{aligned}
    \label{eq:StarkExpansion}
\end{equation}
The mean value of the frequency can be obtained as,
\begin{equation}
    \begin{aligned}
        \langle \nu(p,t) \rangle = \nu_0 &+ \mu(\boldsymbol{E}_0 + p\boldsymbol{F})+\frac{1}{2}\alpha(\boldsymbol{E}_0 + p\boldsymbol{F})^2\\
        &+\frac{1}{2}\alpha \langle \boldsymbol{E}^2_\text{trap}(p,t)\rangle.
    \end{aligned}
    \label{eq:StarkExpansion_mean}
\end{equation}
The center shift can then be derived as,
\begin{equation}
    \begin{aligned}
        \Delta \nu(p) = &(\mu +\alpha\boldsymbol{E}_0)\boldsymbol{F}(p-p_0) +\frac{1}{2}\alpha \boldsymbol{F}^2 (p^2-p_0^2) \\
        &+ \frac{1}{2}\alpha (\omega^2(p)-\omega^2(p_0)),
    \end{aligned}
    \label{eq:StarkExpansion_centerShift}
\end{equation}
where $\omega(p)$ and $\omega(p_0)$ are the standard deviation for the trap field with and without the above-band excitation, respectively. For a finite number of traps, even though they are generated randomly, the total vector sum of their fields at the T center position is unlikely zero, $\boldsymbol{F}\neq0$. We note that if the charge trap distribution is totally symmetric, which leads to $\boldsymbol{F}=0$, we will have $\Delta \nu(p) = \alpha (\omega^2(p)-\omega^2(p_0))/2$. Considering $\alpha<0$ and $\omega^2(p)<\omega^2(p_0)$ (due to the linewidth narrowing), in this case $\Delta \nu(p) > 0$ for all T centers that show linewidth narrowing. This unipolar center shift contradicts with results shown in [Fig.~\ref{fig4:Survey}].

\begin{figure*}[t]
    \centering
    \includegraphics[scale=1]{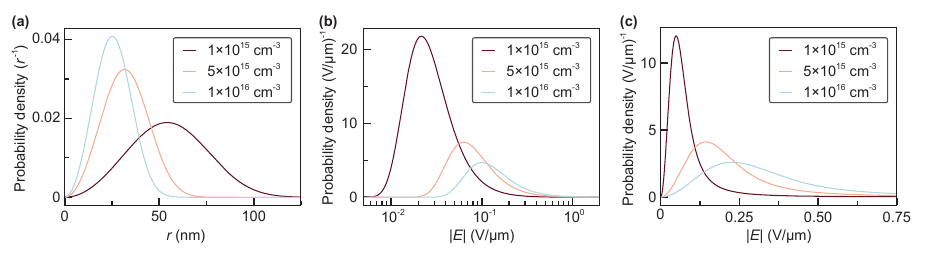}
    \caption{
    \textbf{Trap distance and electric field probability distributions under different trap densities.}
    \textbf{a.} Probability density of finding a trap within distance $r$ from a single T center under an isotropic trap distribution, and \textbf{b.} the corresponding probability density of single-trap fields. 
    % Each color represents a trap density (trap/cm$^3$).
    \textbf{c.} The Holtsmark distribution of total field amplitude evaluated at each trap density.
    % representing the probability distribution of the total field due to the trap environment at a given trap density. 
    }
    \label{fig:sup_trapSpacing}
\end{figure*}

With the narrow and thin bus waveguide structure, surface charges contribute significantly to the built-in field $\boldsymbol{E}_\text{DC}$. These charges mainly include fixed oxide charges ($Q_f$, cm$^{-2}$) inside the native/thermal oxide, and the interface trapped charges ($D_\text{it}$, cm$^{-2}$eV$^{-1}$) due to the silicon dangling bonds (e.g., amphoteric $P_b$ centers). The net charges remaining on the surface can then generate a built-in field inside the waveguide. This field can be roughly estimated assuming 2D surface charges on an infinite plane as $\boldsymbol{E}_\text{DC} = eQ_\text{net}/2\epsilon_\text{Si}$. The finite geometry of the waveguide will have slightly lower and position-dependent fields due to the edge effect. 

In our unpassivated bus waveguide, the two vertical surfaces are generated during the reactive ion etching. The symmetric vertical side walls will generate a zero field at the width center of the waveguide and an increasing field towards the vertical surfaces due to the symmetry. Meanwhile, the top native oxide can hold $Q_f \sim 10^{11} - 10^{12}$ cm$^{-2}$ and the bottom thermal oxide can hold $Q_f \sim 10^{10}$ cm$^{-2}$ \cite{sun1980electron, sharma2015characterizing}. These surface charges can induce spatially-dependent built-in field $|\boldsymbol{E}_\text{DC}| > \, \sim 1$~V/$\mu$m inside the waveguide. We show later in this section that $|\boldsymbol{E}_0| \gg |\boldsymbol{F}|$ and $\alpha (\omega^2(p)-\omega^2(p_0))/2$ is negligible, which then simplifies the center shift to the form shown in Eq.~\ref{eq:centerShift}.

To derive the linewidth, we first write the frequency fluctuation from the mean value,
\begin{equation}
    \label{eq:frequencyFluctuation}
    \begin{aligned} 
        \nu(p,t) - \langle\nu(p,t)\rangle &=
        (\mu+\alpha\boldsymbol{E}_0)\boldsymbol{E}_\text{trap}(p,t)\\
        &+\frac{1}{2}\alpha (\boldsymbol{E}^2_\text{trap}(p,t)- \langle \boldsymbol{E}^2_\text{trap}(p,t)\rangle).
    \end{aligned}
\end{equation}
We can define the variance of the frequency $\nu(p,t)$ as,
\begin{equation}
    \label{eq:frequencyVariance}
    \begin{aligned}
        \text{Var}[\nu(p,t)] &= \langle[\nu(p,t) - \langle\nu(p,t)\rangle]^2\rangle \\
        &=(\mu+\alpha\boldsymbol{E}_0)^2\langle\boldsymbol{E}^2_\text{trap}(p,t)\rangle \\
        &+(\mu+\alpha\boldsymbol{E}_0)\alpha\langle\boldsymbol{E}^3_\text{trap}(p,t)\rangle\\ 
        &+\frac{1}{4}\alpha^2 (\langle\boldsymbol{E}^4_\text{trap}(p,t)\rangle - \langle\boldsymbol{E}^2_\text{trap}(p,t)\rangle^2).
    \end{aligned} 
\end{equation}
Here $\boldsymbol{E}_\text{trap}(p,t)$ originates from the stroboscopically sampled trap fields, which we approximate  as the Holtsmark distribution. From the Holtsmark distribution, however, $\langle\boldsymbol{E}^m_\text{trap}(p,t)\rangle$ diverges for $m \geq 1.5$ due to the heavy tails of the distribution. In physical systems, there exist a high field cut-off typically, which keep the calculation finite. This justifies the standard deviation $\omega(p)$ introduced for $\boldsymbol{E}_\text{trap}(p,t)$ in Eq.~\ref{eq:StarkExpansion_centerShift}. For a specific T center given a trap geometry, the cut-off ($\boldsymbol{E}_\text{cut}$) is mainly affected by the nearest traps, which is then determined by the trap density in a statistical way (Appendix~\ref{appendix:TrapSpacing}).

The Holtsmark distribution of $\boldsymbol{E}_\text{trap}(p,t)$ is symmetric for a vector field in an arbitrary 1D projection, which makes all odd moment expectation values zero ($\langle\boldsymbol{E}^3_\text{trap}(p,t)\rangle = 0$). Considering we typically observe Gaussian-lineshaped T center PLE spectra, we can then obtain the T center FWHM linewidth as, 
\begin{equation}
    \label{eq:fwhm_long}
    \begin{aligned}
        \sigma(p) &= 2\sqrt{2\text{ln}2}  \bigg[(\mu + \alpha \boldsymbol{E}_0)^2\omega^2(p) \\
        &+ \frac{1}{4}\alpha^2(\langle\boldsymbol{E}^4_\text{trap}(p,t)\rangle - \omega^4(p)) \bigg]^{\frac{1}{2}}.
    \end{aligned}
\end{equation}

% Moving on to discuss the linewidth, let us write the variance of the trap field as $\text{Var}[E_\text{trap}(p,t)] = |\omega(p)|^2$. The contribution of the trap field to the linewidth is
% \begin{equation}
%     \label{eq:fwhm_long}
%     \begin{aligned}
%             \sigma(p) &= \\
%             2\sqrt{2\text{ln}|2|} & \sqrt{(\mu + \alpha \boldsymbol{E}_0)^2|\omega(p)|^2 + 1/2\alpha^2 |\omega(p)|^4} ~.
%     \end{aligned}
% \end{equation}

\noindent From the above equation, we first consider the case where the higher order terms are neglected, leading to $\sigma(p) = 2\sqrt{2\text{ln}2}  |\mu + \alpha \boldsymbol{E}_0|\omega(p)$. The typical waveguide-coupled T center has a FWHM linewidth of $\sim$ 1.5 GHz without above-band excitation. With $\boldsymbol{E_0} \sim 1~$V/$\mu$m, $|\mu + \alpha \boldsymbol{E}_0| \sim 10$ GHz/(V/$\mu$m). We can estimate the $\omega(p)\leq\omega(p_0) \sim 0.064$ V/$\mu$m. This justifies $(\mu + \alpha \boldsymbol{E}_0)^2\omega^2(p) \gg (\alpha^2/4)\omega^4(p)$. With reasonable $\boldsymbol{E}_\text{cut}$, one can also have $\langle\boldsymbol{E}^4_\text{trap}(p,t)\rangle \sim \omega^4(p)$, which further simplifies the Eq.~\ref{eq:fwhm_long} and leads to the linewidth narrowing equation shown in Eq.~\ref{eq:linewidthChange}. From the above analysis, it can be shown that $|\boldsymbol{E}_0| \gg \omega(p) \sim |\boldsymbol{F}|$ and $\alpha (\omega^2(p)-\omega^2(p_0))/2 \leq -\alpha\omega^2(p_0)/2 = 0.026$ GHz. These relations justify the omission of all quadratic terms in Eq.~\ref{eq:StarkExpansion_centerShift}.
% 1.5/(2.355)/10

The above modeling also predicts the observed dynamics shown in [Fig.~\ref{fig3:TimeDynamic}]. When both above-band and resonant laser are present, based on the rate equation model, the steady-state $p \sim 1$ under high $P_\text{AB}$. With the pulse sequence used in [Fig.~\ref{fig3:TimeDynamic}(a)(b)] under $P_\text{AB} = 36.1$ $\mu$W, the trap filling dynamics can be described as $p(t_\text{AB}) \approx 1-\exp(-t_\text{AB}/\tau_\text{over})$, where the time constant $1/\tau_\text{over} = \eta^\text{fill}_\text{AB}P_\text{AB}^\text{eff} + \Gamma^\text{fill}_\text{PLE} + \Gamma^\text{detrap}_\text{PLE} + \Gamma^\text{detrap}_\text{dark}$. We note that for the $P_\text{AB} = 36.1$ $\mu$W case, all dynamics finish for $t_\text{AB}\leq500$ ns, we can treat the pulse sequence as a brief resonant-only excitation followed by both excitations for $t_\text{AB}$ duration. Considering the large depletion length from the surface charges, the trap filling at the beginning of the sequence is close to zero.
Based on Eq.~\ref{eq:centerShift} and Eq.~\ref{eq:linewidthChange}, the center shift and linewidth narrowing are $\Delta\nu(t_\text{AB}) \approx C_\text{f}[1-\exp(-t_\text{AB}/\tau_\text{AB})]$ and $\Delta \sigma(t_\text{AB}) = C_\text{L}[\exp(-t_\text{AB}/\tau_\text{AB})^{2/3} - 1]$, which indicates the factor 1.5 times slower dynamics for the linewidth narrowing than the center shift. In the case of $P_\text{AB} = 2.1$ $\mu$W, there is no simple form of $p(t)$, necessitating the numerical model to describe the dynamics.

With the pulse sequence used in [Fig.~\ref{fig3:TimeDynamic}(e)(f)], the trap filling dynamics can be described as $p(t_\text{PLE}) \approx \exp(-t_\text{PLE}/\tau_\text{PLE})$ (Appendix~\ref{appendix:timeConstantVsPower}). Based on Eq.~\ref{eq:centerShift} and Eq.~\ref{eq:linewidthChange}, the center shift and linewidth narrowing are $\Delta\nu(t_\text{PLE}) \approx C_\text{f}\exp(-t_\text{PLE}/\tau_\text{PLE})$ and $\Delta \sigma(t_\text{PLE}) = C_\text{L}[(1-\exp(-t_\text{PLE}/\tau_\text{PLE}))^{2/3} - 1]$, respectively. We note that the linewidth change dynamics
% depending on the resonant laser pulse width $t_\text{PLE}$ 
has an effective time constant of $0.7\tau_\text{PLE}$ using the typical $1/e$ decay threshold for a standard single exponential function. This ratio roughly matches with the observed time constants for center shift and linewidth narrowing dynamics as shown in [Fig.~\ref{fig3:TimeDynamic}(e)(f)] and [Fig.~\ref{figSI:supG_timeConstantVsPower}].

\section{TRAP FIELD AND SPACING}
\label{appendix:TrapSpacing}

The nature of the single T center's local charge environment determines its spectral properties and response to the above-band excitation. In particular, the traps closest to the T center can have an out-sized impact on the Stark shift, and lead to an effectively anisotropic local field environment. The probability distribution of finding the closest trap to be at distance $r$ from the T center [Fig.~\ref{fig:sup_trapSpacing}(a)] can be evaluated \cite{johnston2024cavity} by, 
\begin{equation}
    \label{eq:trap_spacing}
    W(r) = 4\pi \rho^2 \text{exp}\left(-\frac{4\pi\rho}{3} r^3\right),
\end{equation}
where $\rho$ is the trap density. Correspondingly, the field amplitude distribution from a single trap [Fig.~\ref{fig:sup_trapSpacing}(b)] can be calculated via a variable transformation from Eq.~\ref{eq:trap_spacing}. 
% and the probability distribution of the field due to the closest trap \cite{fujimoto2004plasma}.  
However, the field that the T center experiences is better described by the vector sum of all trap fields. Given a trap density ($\rho$), Holtsmark model describes the distribution of total field from all traps under a large number of different spatial geometries [Fig.~\ref{fig:sup_trapSpacing}(c)], and can be obtained as \cite{holtsmark1919verbreiterung, fujimoto2004plasma},
\begin{equation}
    \label{eq:HoltsmarkDist}
    W(\beta) = \frac{2}{\pi\beta} \int_0^\infty x\sin(x)\exp\left[-(x/\beta)^{3/2} \right]dx,
\end{equation}
where $\beta = |E|/E_\text{c}$ normalizes the field amplitude to a characteristic Holtsmark field \cite{holtsmark1919verbreiterung},
\begin{equation}
    \label{eq:HoltsmarkEc}
    E_\text{c}=2.6\left(\frac{e}{4\pi\epsilon_0\epsilon_r}\right)\rho^{2/3}.
\end{equation}
% To gain a more complete picture of the local electric field value, the Holtsmark distribution is employed, which describes the distribution of total field values generated by an infinite and isotropic distribution of fluctuating charges . 
% Fig \ref{fig:sup_trapSpacing}(c.) shows the Holtsmark distribution evaluated at three densities, the long tail of the distribution is representative of the dominant effect that the traps closest to the emitter have on the total field. 
Considering a noise-field estimate from Appendix~\ref{appendix:modeling_details} of $\omega(p_0) \approx 0.064$ V/$\mu$m, a Holtsmark distribution with a corresponding field requires $\rho\sim8\times10^{15} $ cm$^{-3}$. For~this estimation we match the linewidth-corresponding field $2\sqrt{2\ln2}\omega(p_0)$ with Holtsmark's relation of $\sigma/c= 1.25E_c$ for linewidth ($\sigma$) of an emitter with a linear Stark coefficient $c$ in the presence of Holtsmark fields \cite{holtsmark1919verbreiterung}.
% (e_charge_C/(4*np.pi*epsilon0_Fm*epsilon_r)) = 1.172e-10 # C
% Then, FWHM = 1.25*2.603*1.172e-10*rho**(2/3)
% FWHM = 2*sqrt(2*ln(2))*0.064 = 0.15071 V/um
% 1e-6*(0.1507e6/3.8134e-10)^(3/2) = 7.856E15 trap/cm3
% Thus, the distributions in Fig. \ref{fig:sup_trapSpacing} can give a sense of the value of the fixed trap field, $\boldsymbol{F}$.

\section{T CENTER LINEWIDTH NARROWING STATISTICS} \label{appendix:statistics}
\begin{figure}
        \centering
        \includegraphics[scale=1]{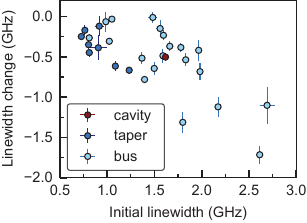}
        \caption{
        \textbf{Survey of T center linewidth narrowing.} Statistics of T center maximum linewidth narrowing against its initial linewidth without above-band excitation. We also add data for T centers in the taper waveguide and cavity.     
        }
        \label{fig:sup_statisitcs}
\end{figure}

From the discussion in Sec.\ref{section:modeling}, the largest linewidth narrowing for a specific T center should depend roughly linearly on its initial linewidth without the above-band excitation. This is shown as we change the $x$-axis in [Fig.~\ref{fig4:Survey}] to the initial linewidth [Fig.~\ref{fig:sup_statisitcs}]. Saturated linewidth change and center shift for T centers in the taper waveguide and cavity behave similarly as T centers in the bus waveguide. However, T centers in the wider tapered waveguide typically show narrower initial linewidth. For cavity-coupled single T center, we intentionally detune the cavity to minimize its modulation of the T center spectrum during center shift due to the above-band excitation.

% Appendix.\ref{appendix:modeling_details}, the T center with larger optical linewidth suffers from larger built-in field due to the impurities in silicon. Consequently, the change in linewidth induced by filling the fluctuating charge traps is also scaled up. Fig.~\ref{fig:sup_statisitcs} shows the statistics of such behavior, where the change in linewidth and original linewidth has a strong negative correlation, supporting the modeling shown in early sectinos.

\newpage
\bibliography{reference.bib}        

\end{document}